\documentclass[aps,pra,twocolumn,amsmath,amssymb,floatfix,eqsecnum,showpacs,nofootinbib]{revtex4}
\usepackage{amsmath}
\usepackage{graphicx}
\usepackage{latexsym}
\usepackage{amsmath,amssymb}
\usepackage{bm} 
\usepackage{color}
\usepackage{stackrel}
\usepackage{accents}

\DeclareMathOperator{\sgn}{sgn}

\newcommand{\ord}[1]{\bm{\mathit{O}}\left(#1\right)}

\newcommand{\vex}[1]{\bm{\mathrm{#1}}}

\newcommand{\sss}[1]{\scriptscriptstyle{#1}}
\newcommand{\msf}[1]{\mathsf{#1}}

\newcommand{\ket}[1]{\left| {#1} \right\rangle}

\newcommand{\pup}[1]{\scriptscriptstyle{({#1})}}

\DeclareMathOperator{\re}{Re}
\DeclareMathOperator{\im}{Im}

\newcommand{\bsub}{\begin{subequations}}
\newcommand{\esub}{\end{subequations}}

\newcommand{\e}{\varepsilon}

\newcommand{\emax}{\varepsilon_\Lambda}
\newcommand{\Dasy}{\Delta_{\sss{\infty}}}
\newcommand{\masy}{\mu_{\sss{\infty}}}

\newcommand{\Do}{\Delta_{\sss{0}}}

\newcommand{\Dqcp}{\Delta_{\msf{QCP}}}

\newcommand{\intem}{\int_0^{\emax} d \e\,}
\newcommand{\Vmin}{V_{\msf{min}}}

\newcommand{\G}{\mathcal{G}}

\newcommand{\Di}{\Delta_{\sss{0}}^{{\sss{(}}i{\sss{)}}}}
\newcommand{\mui}{\mu_{\sss{0}}^{{\sss{(}}i{\sss{)}}}}

\newcommand{\Df}{\Delta_{\sss{0}}^{{\sss{(}}f{\sss{)}}}}
\newcommand{\muf}{\mu_{\sss{0}}^{{\sss{(}}f{\sss{)}}}}

\newcommand{\dos}{\nu_0}

\newcommand{\Rhot}{\msf{R}}

\newcommand{\muco}{\msf{m}}

\DeclareMathOperator{\cn}{cn}

\newcommand{\rr}{\mathfrak{r}}
\newcommand{\ii}{\mathfrak{i}}

\newcommand{\wRhot}{\widetilde{\mathsf{R}}}

\newcommand{\Emin}{E_{\msf{min}}}

\newcommand{\erf}{\mathcal{E}_{2,1}}
\newcommand{\kb}{\vex{k}}

\begin{document}

\title{
	Spectroscopic
	probes of 
	isolated
	nonequilibrium quantum matter: \\
	Quantum quenches, Floquet states, and distribution functions
}
\author{Yunxiang Liao} \email{yliao@rice.edu}
\affiliation{Department of Physics and Astronomy, Rice University, Houston, Texas 77005, USA}
\author{Matthew~S.~Foster}
\affiliation{Department of Physics and Astronomy, Rice University, Houston, Texas 77005, USA}

\date{\today\\}

\begin{abstract}
We investigate radio-frequency (rf) spectroscopy, metal-to-superconductor tunneling, and angle-resolved photoemission spectroscopy (ARPES) as probes of isolated 
out-of-equilibrium quantum systems, and examine the crucial role played by the nonequilibrium distribution function. 
As an example, we focus on the induced topological time-periodic (Floquet) phase in a two-dimensional $p + i p$ superfluid, following 
an instantaneous quench of the coupling strength. The post-quench Cooper pairs 
occupy a linear 
combination of ``ground'' and ``excited'' Floquet states, with coefficients determined by the distribution function.
While the Floquet bandstructure exhibits a single avoided crossing relative to the equilibrium case, 
the distribution function shows a \emph{population inversion} of the Floquet bands at low energies. 
For a realization in ultracold atoms, these two features compensate, producing a bulk average rf signal 
that is well-captured by a quasi-equilibrium approximation. In particular, the rf spectrum shows a robust gap.
The single crossing occurs because the quench-induced Floquet phase belongs to a particular
class of soliton dynamics for the BCS equation. 
The population inversion is a consequence of this, and ensures 
the conservation of the pseudospin winding number.
As a comparison, we compute the rf signal when only the lower Floquet band
is occupied; in this case, the gap disappears for strong quenches. 
The tunneling signal in a solid state realization is ignorant of the distribution function, and
can show wildly different behaviors.
We also examine rf, tunneling, and ARPES for weak quenches, such that the resulting topological steady-state is characterized
by a constant nonequilibrium order parameter. 
In a system with a boundary, tunneling reveals the Majorana edge states. 
However, the local rf signal due to the edge states is suppressed by a factor of the inverse system size, and is 
spatially deconfined throughout the bulk of the sample.
\end{abstract}

\maketitle

\tableofcontents

\section{Introduction \label{Sec: Intro}}

\subsection{Idea and overview of results}

In recent years, there 
have	
been numerous studies on Floquet systems as they possess interesting topological 
phases absent in static systems 
\cite{OkaAoki09,Lindner11,Kitagawa11,Gu11,Rudner13,Kundu14,Usaj14,SHORT}. 
Most of these 
are induced by external driving. For example, a Floquet topological insulator 
may arise 
when graphene is 
irradiated	
by 
circularly polarized light \cite{OkaAoki09,Kitagawa11,Gu11,Kundu14,Usaj14}. However, 
a	
topological 
Floquet phase can also be created in an isolated system 
such as a BCS superfluid
following an instantaneous quench of 
the	
interaction strength 
\cite{SHORT,LONG,DongPu14}.	
The periodic 
modulation of the order parameter 
is self-generated by dynamics,
and arises due to the single unstable mode 
associated to the BCS instability of the normal state
\cite{S-wave1,YuzbashyanAltshuler06,BarankovLevitov06,BarankovLevitov06-II}.

In either of these two cases, the Floquet (quasienergy) band structure and the distribution function 
(nonequilibrium occupation number) are equally important. Much of the previous work focused on the 
connection between the band structure and topological properties \cite{Lindner11,Rudner13}, but did not take into 
account the distribution function which might be essential to 
determine
experimental observables. 
The problem of 	
irradiated graphene
has been considered mainly in the Landauer-B\"uttiker formalism, wherein 
the mode occupation is fixed by ideal leads \cite{Kitagawa11,Gu11,Kundu14,Usaj14}.
Even in that simplified setting, however, the 
relationship 
between the topology and Floquet edge states
to measured transport coefficients is complicated, and the latter are not generically quantized \cite{Gu11,Kundu14,Usaj14}.
More recent works have considered Floquet systems that are isolated or weakly coupled to the environment 
\cite{Chamon13,Mitra14,DAlessioRigol14}.
If external driving is applied to a system initially in its ground state, the particles will populate
several Floquet bands, leading to a nontrivial distribution function.

To 
better	
understand the connection between the distribution function and 
observables, in this work	
we study the radio-frequency (rf) spectrum \cite{rfswave} of 
an isolated
topological Floquet system. 
We consider the topological $p + i p$ Floquet superfluid 
induced by an instantaneous quench of interaction strength
in a 2D system of spinless fermions \cite{SHORT,LONG}.	
The long-time asymptotic steady state occupies two Floquet bands related by particle-hole 
symmetry. We determine the distribution function as well as the explicit BCS dynamics of the steady state, and derive the 
expression 
for the 
rf spectrum. This out-of-equilibrium system 
could be	
realized in 
an	
ultracold gas of fermionic atoms interacting through 
a	
$p$-wave Feshbach resonance \cite{SHORT,LONG}. 

Assuming it can also be realized in solid, we discuss the possibility of probing this nonequilibrium 
system by normal metal to superconductor tunneling \cite{Schrieffer}, whose setup closely parallels that of rf. 
Finally we consider momentum-resolved spectra, as could be measured using angle-resolved photoemission spectroscopy (ARPES) \cite{GedikARPES}.
Recent THz pump-probe experiments 
\cite{Shimano1,Shimano2}
have shown that order parameter (``Higgs mode'')  quench dynamics can be 
induced in low temperature superconductors on ultrashort timescales.
Our results apply in a transient window over which the time-periodic phase 
can be stabilized, before pair-breaking or other processes destroy it \cite{BarankovLevitov06-II,SHORT}.	

We show that the bulk rf and tunneling spectra are in general very different, and this
is tied to the nonequilibrium distribution function. 
The rf and tunneling spectra are both one-particle observables, but there is a crucial distinction \cite{rf-Tun-SecondDist}. 
In an ultracold gas, rf radiation can induce internal transitions in the atoms between states that 
participate in pairing and states that do not. The rf signal depends upon the occupation of both the paired and unpaired states. 
By contrast, tunneling from an idealized metal tip gives a current that is sensitive to the distribution function 
in the superconductor only on energy scales of order $\mathcal{E}_F$, the Fermi energy in the tip. 
The tunneling conductance at small bias is completely determined by the retarded single particle Green's function
in the Floquet phase. The same quantity determines transport in Floquet scattering theory \cite{Kitagawa11},
and signals the absence or presence of edge states in a system with a boundary \cite{LONG,SHORT,Gurarie11}; however, it 
does not encode the occupation of bulk or edge states. 

In our quenched-induced topological superfluid, we show that there is exactly one avoided
crossing in the Floquet bandstructure. At the same time, there is a \emph{population inversion} in
the occupation of these bands, so that the ``upper'' (``lower'') quasienergy band is occupied
at low (high) energies. These two features compensate in the bulk rf prediction, so that
the average signal is well-approximated by that of a quasi-equilibrium $p+ip$ superfluid.
In particular, the spectrum exhibits a robust gap. Deviations from a quasi-equilibrium picture
appear as satellite coherence peaks, which have been studied previously \cite{rfswave}.
To compare, we also consider
the rf signal that obtains by populating only the lower Floquet band. In this case the
gap disappears for sufficiently strong quenches that induce large variations in the order parameter
(i.e., strong driving). 
The tunneling signal is ignorant of the distribution function, and also depends
sensitively on the quench. For strong quenches there is no gap in the bulk tunneling spectrum. 
A quench-induced population inversion in a single band Floquet system was studied in  
\cite{Tsuji2011}. 

Unlike toy models sometimes considered in the literature, the drive frequency in our quench-induced
Floquet system is much smaller than the bandwidth. This implies that there is significant 
folding of the unperturbed spectrum across the quasienergy zone. Although this situation should
be the norm rather than the exception experimentally \cite{OkaAoki09}, it is usually more
complicated to understand than the opposite, high frequency limit \cite{Kitagawa11,Rudner13}. In particular, one
typically expects small gaps to open every time the spectrum crosses the zone, and this can
complicate predictions for experiment \cite{Gu11}. 

By contrast, the presence of a single avoided band crossing in our Floquet bandstructure is a consequence 
of the BCS dynamics that generate it. In particular, the crossing arises because the
quench-induced Floquet phase belongs to a particular class of soliton dynamics for the
integrable BCS equations \cite{YuzbashyanAltshuler06,BarankovLevitov06}.
The population inversion is then a topological consequence of this, since the
texture of the instantaneous Anderson pseudospin description of the BCS state is
conserved by the dynamics \cite{LONG,SHORT}. 
In other isolated 
topological Floquet systems, a population inversion is also expected for an 
odd number of avoided crossings. 

We also consider the 
rf, tunneling, and ARPES
signals that result from weaker quenches
in the $p+ip$ superfluid, such that the order parameter asymptotes to a nonequilibrium constant 
value. In addition to bulk spectra, we consider a semi-infinite geometry with an edge. 
We show that while tunneling reveals the Majorana edge states, local rf spectroscopy
is ill-suited to find them. This is because the signal is suppressed by a factor of the
system size, and is not spatially localized to the edge of the cloud.

\subsection{Review: Quench-induced topological Floquet phase in a p+ip superfluid}

In this subsection, we briefly review the quench induced nonequilibrium state 
\cite{SHORT,LONG} whose rf spectrum is investigated.
The system is governed by the BCS Hamiltonian,
\begin{align}\label{eq:H0}
	H =
	\sum_{\kb} \frac{k^2}{2}
	c^{\dagger}_{\kb} c_{\kb}
    -2 G \sum_{\kb,\vex{q}}'
    \kb \cdot \vex{q}
	\,
	c^{\dagger}_{\kb} c^{\dagger}_{-\kb} 
	c_{-\vex{q}} c_{\vex{q}},
\end{align}
where $c_{\kb}$ annihilates 
a
fermion with momentum 
$\vex{k} = \{k^x,k^y\}$ 
and mass $m=1$. The prime in $\sum'$ indicates that the summation runs over momenta with nonnegative $y$-components, i.e. $k^y \geq 0$. $G>0$ is the interaction strength.

This Hamiltonian [Eq.~(\ref{eq:H0})] has a $p+ip$ ground state whose order parameter assumes the form
\begin{align}
	\Delta(\kb) & \equiv
	-2G
	\sum_{\vex{q}}'
	\kb \cdot \vex{q} 
	\langle
	c_{-\vex{q}} c_{\vex{q}}
	\rangle
	\nonumber\\
	&=\Delta_0(k^x-ik^y)
	 =\Delta_0 k e^{-i\phi_k},
\end{align}
where $\phi_k$ is the polar angle of 2D vector $\kb$. $\Delta_0$ is the order parameter amplitude. 
The quasiparticle energy of such 
a
$p+ip$ paired state is
\begin{align}\label{Ek}
	E_k=
	\textstyle{
	\sqrt{
	\left(\frac{k^2}{2}-\mu \right)^2
	+k^2 \Delta_0^2
	 }}.
\end{align}
Its spectrum is fully gapped as long as the chemical potential $\mu \neq 0$.
The critical point $\mu=0$ separates two distinct topological phases 
\textemdash  
the weak pairing BCS phase ($\mu>0$) and the strong pairing BEC phase ($\mu<0$). 
The BCS phase is topologically nontrivial, i.e.\ if the system in this phase possesses 
a boundary, the Majorana edge state would appear.
The squared order parameter $(\Delta_0)^2$ and chemical potential $\mu$ carry units
of density; the fixed particle density $n$ sets the natural scale. All dimensionful quantities
for a quench can be expressed in units of $n$, with at most logarithmic dependence upon
an ultraviolet
energy
cutoff $\Lambda$ \cite{LONG}.

The system is initially prepared in the $p+ip$ ground state of 
the
pre-quench Hamiltonian with interaction strength $G_i$. Then the interaction strength is suddenly 
quenched to a different value $G_f$. After the quench, the system evolves as a superposition of 
many body eigenstates of 
the
post-quench Hamiltonian and acquires a steady state as $t\rightarrow \infty$. 

\begin{figure}
\includegraphics[width=0.4\textwidth]{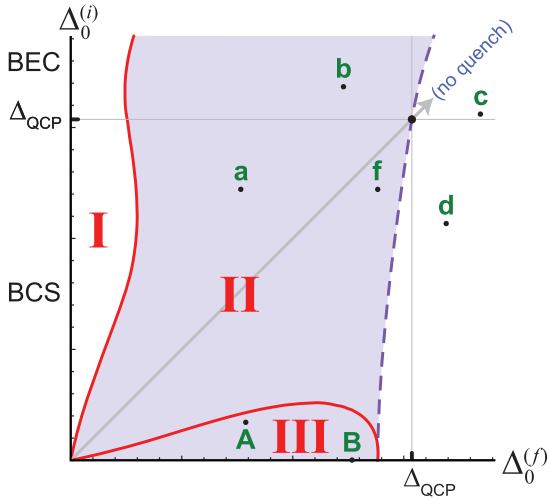}
\caption{
(color online).
Quench phase diagram \cite{footnote--Phase Diagram}	
showing 3 distinct dynamical phases characterizing the out-of-equilibrium states induced by sudden quench of the interaction strength
in a 2D $p+ip$ fermion superfluid \cite{LONG,SHORT}. 
The order parameter amplitude $\Dasy(t)$ either vanishes, converges to a nonzero constant, or oscillates persistently for quenches in phase I, II, 
or III, respectively. 
The vertical and horizontal axes show $\Di$ and $\Df$, the order parameter amplitudes associated to the 
ground states of the pre- and post-quench Hamiltonians.
$\Dqcp$ indicates the critical point where the ground state chemical potential vanishes. It separates the 
topologically nontrivial BCS and trivial BEC 
phases in equilibrium.
The purple dashed curve is its nonequilibrium extension. 
Each point to the left (right) of this line in 
II 
represents a topologically nontrivial (trivial) nonequilibrium state with positive (negative) $\masy$.
Phase III is a quench-induced Floquet topological superfluid state, which is our main focus here. 
The labeled points are particular quenches discussed in the text.
The amplitudes ${\Delta_{\sss{0}}^{{\sss{(}}i,f{\sss{)}}}}$ and $\Dqcp$ are measured in units
of the inverse length $\sqrt{n}$, where $n$ is the fixed particle density \cite{LONG}.
}
\label{fig:diagram}
\end{figure}

In general, the 
asymptotic long-time
evolution of the order parameter amplitude 
can be expressed as 
\begin{align}\label{DasyDef}
	\Delta(t) =\Delta_{\infty} (t) \, e^{-2 i \masy t},
\end{align}
where $\Delta_{\infty} (t)$ 
could be 0 (phase I), a nonzero constant (phase II) or an oscillating function with time period $T$ (phase III), 
depending on the strength and direction of the quench
\cite{BarankovLevitov06,DzeroYuzbashyan06,LONG,SHORT}.
In phase II, the real constant $\masy$ 
plays the role of a nonequilibrium chemical potential and determines
the topological properties of the quench induced state. It is topologically nontrivial (trivial) 
when $\masy>0$ ($\masy<0$). 
The out-of-equilibrium phase diagram is shown in Fig.~\ref{fig:diagram} 
where we use $\Do^{(i)}$ and $\Do^{(f)}$, the order parameter amplitudes of pre- and post-quench Hamiltonians' 
ground states, to indicate the coordinates of the quench \cite{LONG,SHORT}.

The main interest in this work is phase III, which can be understood
as a quench-induced Floquet topological superfluid phase \cite{SHORT}. 
For both phases II and III, we use the term ``topological'' to denote
the presence or absence of Majorana edge modes in a sample with a boundary,
as encoded in the single particle retarded Green's function \cite{LONG,SHORT,Gurarie11}. 
This notion of topology relates to the spectrum of excitations that can be 
reached from the quench-induced nonequilibrium state by the application of
a weak spectral probe, i.e.\ a frequency-dependent measurement. 
There is a different notion of topology that describes the many-body wavefunction itself, which is here always a BCS state.
This is the pseudospin winding number that can be observed by an equal-time measurement,
such as time-of-flight, and which cannot change following a quench.
The conserved pseudospin winding does not determine whether or not edge states
will appear; it instead encodes the occupation of the states \cite{LONG,SHORT}.
Invariance of the winding number of the state was also noted in 
externally driven Floquet \cite{DAlessioRigol14} and quenched topological \cite{Caio15} models.

In the following, we focus on weak-to-strong quenches within phase III where the order parameter amplitude 
is given by a Jacobi elliptic function \cite{LONG}. The constant $\Delta_{\infty}$ in phase II can be considered 
as a trivial case with 
vanishingly small oscillation amplitude.

\subsection{Floquet states and distribution function}

The order parameter amplitude $\Dasy(t)$ is 
determined by the self-consistent BCS dynamics \cite{YuzbashLax1,YuzbashLax2,YuzbashyanAltshuler06,BarankovLevitov06,DzeroYuzbashyan06,LONG}.
The dynamics of individual Cooper pairs or (pairs of quasiparticle excitations) can then be obtained by
solving the effective mean field Hamiltonian,
\begin{align}\label{eq:BdGH}
	H_{\msf{BdG}}=\sum_{\kb}'
	\left\lbrace 
	\begin{aligned}
		& {\textstyle{\left(\frac{k^2}{2}-\masy\right)}} (c^{\dagger}_{\kb} c_{\kb}+c^{\dagger}_{-\kb} c_{-\kb})
		\\
		& +k \Dasy(t) \, c^{\dagger}_{\kb}c^{\dagger}_{-\kb}
		+k \Dasy^{*}(t) \, c_{-\kb}c_{\kb}
	\end{aligned}
	\right\rbrace\!. 
\end{align}
Here 
we have absorbed the polar phase of $\Delta(\vex{k})$ into the fermion operators via
$c_{-\kb}c_{\kb} \rightarrow  e^{-i \phi_{\kb}-i 2\masy t} c_{-\kb}c_{\kb}$. 
This transformation also boosts into the ``rotating frame,'' so that $\masy$ is transferred
from the order parameter amplitude to the kinetic term.

The corresponding many body wavefunction is given by the BCS product
state
\begin{align}\label{eq:BCSproduct}
	|\Psi(t)\rangle
	=
	\prod_{\kb}'
	\left[ 
	u_{\kb}(t)+v_{\kb}(t)
	c_{\kb}^{\dagger}
	c_{-\kb}^{\dagger}
	\right]
	|0\rangle,
\end{align}
where the time-dependent coherence factors follow the Bogoliubov-de Gennes 
(BdG)
equation
\begin{align}\label{eq:BdGeq}
	i \frac{d }{d t}
	\begin{bmatrix}
	u_{\vex{k}}(t) \\
	v_{\vex{k}}(t)
	\end{bmatrix}
	=&\,
	\begin{bmatrix}
	 - \frac{k^2}{2}+\masy
	 & k \Dasy^*(t)
	 \\ 
	 k \Dasy(t)
	 & \frac{k^2}{2} -\masy
	\end{bmatrix}
	\begin{bmatrix}
	u_{\vex{k}}(t) \\
	v_{\vex{k}}(t)
	\end{bmatrix}.
\end{align}

When $\Dasy(t)$ is periodic in time 
as in phase III \cite{SHORT} with
$\Dasy(t+T)=\Dasy(t)$, Eq.~(\ref{eq:BdGeq}) has a pair 
of 
Floquet 
solutions.
These are
states taking the form 
$|\Psi^{(F)}(t)\rangle=|\Phi^{(F)}(t)\rangle e^{i E^{(F)} t}$, 
where 
$|\Phi^{(F)}(t)\rangle$ shares the same period $T$ with $\Dasy(t)$, 
$|\Phi^{(F)}(t+T)\rangle=|\Phi^{(F)}(t)\rangle$. 
The
quasi-energy $E^{(F)}$ is 
defined
up to integer multiples of 
the
frequency $\Omega=2\pi/T$. 
A
Floquet state 
is a solution
subject to the 
special
initial condition that $|\Psi^{(F)}(0)\rangle$ is an eigenstate of time evolution operator 
$U(T)$
over one period:
$U(T)|\Psi^{(F)}(0)\rangle=e^{i E^{(F)} T}|\Psi^{(F)}(0)\rangle$.

We will always assume a particular type of physical initial condition for the many-fermion system, which is
the $p+ip$ ground state of 
the
pre-quench Hamiltonian. 
The coherence factors in the asymptotic steady state are then given by 
a superposition of two Floquet states, both of which solve the BdG equation Eq.~(\ref{eq:BdGeq}), that is
\begin{align}\label{eq:BdGsol}
\begin{aligned}
	\begin{bmatrix}
	  u_{\kb} (t)
	  \\
	  v_{\kb} (t)
	\end{bmatrix}  
	 =&\,
	 \sqrt{
	   \frac{1- \gamma_{\kb}}{2}
	      }
	\begin{bmatrix}
	  u_{{\kb}} ^{(F)} (t)
	  \\
	  v_{{\kb}} ^{(F)} (t)
	\end{bmatrix}
     e^{+ i E^{(F)}_{\kb} t}
     \\
     &+\,
     \sqrt{
     	\frac{1+ \gamma_{\kb}}{2}
     	  }
     \begin{bmatrix}
       v_{{\kb}} ^{*(F)} (t)
       \\
       -u_{{\kb}}^{*(F)} (t)
     \end{bmatrix} 
     e^{- i E^{(F)}_{\kb} t + i \Gamma_{\kb}}.
\end{aligned}
\end{align}
These two Floquet states are related by particle-hole symmetry, and the relative coefficients are 
set
by the 
``Cooper pair distribution function''
$\gamma_{\kb}$
($-1 \leq \gamma_{\kb} \leq 1$), 
which determines the nonequilibrium occupation number
$f_{\kb}^{\pup{\mathsf{QP}}}$
of 
the	
Floquet bands \cite{LONG,rfswave}.
These are related via
\begin{align}\label{eq:fqpDef}
	f_{\kb}^{\pup{\mathsf{QP}}} = {\textstyle{\frac{1}{2}}}\left(1 + \gamma_{\kb}\right),
\end{align}
so that $\gamma_{\kb} = -1$ ($+1$) corresponds to the absence (presence) of a pair of excited 
quasiparticles
occupying states $\pm \kb$. 
This is different from the occupation of the bare fermions 
$\langle(c^{\dagger}_{\kb} c_{\kb}+c^{\dagger}_{-\kb} c_{-\kb})\rangle$.

We will demonstrate in the following sections that the distribution function possesses topological 
information of the pre- and post- quench states. It can be extracted from the rf spectroscopy but not 
from the superconductor-normal metal tunneling experiment.

\subsection{Outline}

This paper is organized as follows. In Sec.~\ref{Sec:Setup}, we 
derive general expressions for 
rf spectroscopy, tunneling 
amplitudes, and ARPES
employed to probe the quench-induced out-of-equilibrium system. 
In Sec.~\ref{Sec:II}, 
as a warm-up 
we compute the rf, tunneling, and ARPES spectra associated with several 
quenches in phase II with various topological properties. 
We 
examine 
the possibilities of extracting 
the 
topological character of both pre- and post- quench states by analyzing the rf spectrum. 
In Sec.~\ref{Sec:III}, we 
turn to the analysis of the Floquet phase III. We
study thoroughly two representative phase III quenches: one 
located
in the vicinity of 
the
phase II--III boundary, while the other lies deep 
in the phase III with weak initial interaction strength. 
The former (latter) is characterized by small harmonic 
(large anharmonic) amplitude modulations in $\Dasy(t)$.
We present and discuss the results for the rf, tunneling, and ARPES spectra,
and we also discuss the connections between the Floquet bandstructure
and the integrable BCS dynamics.
Finally, in Sec.~\ref{Sec:Inhomogeneous} we compute the local rf signal of 
a phase II
post-quench system with a boundary, looking for signatures of Majorana edge 
states.
We give our conclusion in Sec.~\ref{Sec: End}. 

Expressions for rf and tunneling harmonics relevant
to probing the Floquet phase are relegated to
Appendix \ref{Sec:App0}.
The detailed derivations of the 
explicit
Floquet state 
wavefunctions
as well as the distribution function 
associated with phase III 
quenches
are given in Appendix \ref{Sec:App1}.

\section{RF, tunneling spectroscopy, and ARPES \label{Sec:Setup}}

\subsection{Hamiltonian and current}

In
an
rf experiment, we assume
that
the system under study is realized in 
an ultracold fermion gas that
possesses two 
relevant
hyperfine states: $|1\rangle$ which participates in pairing, 
and $|2\rangle$ which does not.

The Hamiltonian 
for atoms in the non-pairing state $|2\rangle$
is given by
\begin{align}\label{eq:rfH0d}
\begin{aligned}
	H_{0}^{(d)} =
    \sum_{\vex{k}}
	  {\textstyle{\left(\frac{k^2}{2} + \erf\right)}}
	  d_{\kb}^{\dagger}d_{\kb},
\end{aligned}
\end{align}
where $\erf$ denotes the atomic transition energy between the states
$|2\rangle$ and  $|1\rangle$.
This energy separation is typically much larger than any relevant to the 
many-body dynamics.
 $d_{\kb}$ annihilates 
an atom
in state $|2\rangle$ with momentum $\kb$.
In a $p$-wave paired superfluid, the rf radiation induces a transition 
in an atom with momentum $\vex{k}$ from state $|1\rangle$ to $|2\rangle$;
an unpaired state $|1\rangle$  atom with momentum $-\vex{k}$ is left behind. 
This process is described by the transition (coupling) Hamiltonian
\begin{align}\label{eq:rfHT}
\begin{aligned}
	H_{T} =
	\mathcal{T}
    \sum_{\vex{k}}
    \left[ 
	  e^{i \omega_L t}
	  c_{\kb}^{\dagger}d_{\kb}
	  +
	  e^{-i \omega_L t}
	  d_{\kb}^{\dagger}c_{\kb}
	  \right],
\end{aligned}
\end{align}
where $\omega_L$ is the frequency of the rf field.

We will consider two observables: the 
local 
rf current 
$I(\vex{r}_0) \equiv \langle d n_d(\vex{r}_0) / d t \rangle$
and the
global current 
$I \equiv \langle d N_d/d t \rangle$.
Here $n_d(\vex{r}_0)$ is the spatial density distribution of $|2\rangle$ atoms at position $\vex{r}_0$,
\begin{align}\label{eq:rfNd}
\begin{aligned}
	n_{d}(\vex{r}_0) 
	\equiv
	d^{\dagger}(\vex{r}_0) d(\vex{r}_0)
	=
    \sum_{\kb_1,\kb_2}
    \left[ 
      d^{\dagger}_{\kb_1}
      d_{\kb_2}
	  e^{i(\kb_1- \kb_2)\vex{r}_0}
	  \right], 
\end{aligned}
\end{align}
while $N_d$ is the total number of atoms in state $|2\rangle$
\begin{align}\label{eq:rfNdT}
\begin{aligned}
	N_{d}=
	 \sum_{\kb}
    d^{\dagger}_{\kb} d_{\kb}.  
\end{aligned}
\end{align}

Aside from the rf, we also consider normal metal to superconductor tunneling, assuming 
that
the same post-quench 
asymptotic steady state can be 
hypothetically
realized in 
a superconductor. 
The 
quenched
condensate is brought into contact with a normal metal tip, whose Hamiltonian is given by
\begin{align}\label{eq:TUNHd}
\begin{aligned}
	H_{0}^{(d)} =
    \sum_{\vex{k}}
	  {\textstyle{\left(\frac{k^2}{2} + V\right)}}
	  d_{\vex{k}}^{\dagger}d_{\vex{k}}.	  
\end{aligned}
\end{align}
Here the tunneling bias $V$ is the energy of the tip relative to the condensate. 
We assume that the tip 
is ideal, characterized by
a constant density of states $\dos$, and
that prior to contact 
only states with momentum $k \leq q_F$ are occupied.
The Hamiltonian
\begin{align}\label{eq:TUNHT}
\begin{aligned}
	H_{T} =&
	\mathcal{T}
	  \left[ 
      c^{\dagger}(\vex{r}_0) d(\vex{r}_0)
	  +
	  d^{\dagger}(\vex{r}_0) c(\vex{r}_0)
	  \right] 
	  \\
	 =&
	 \mathcal{T}
	  \sum_{\kb,\vex{q}}
	  \left[ 
	    c^{\dagger}_{\kb}
	    d_{\vex{q}}
	 	e^{i(\kb- \vex{q})\vex{r}_0}
	 	+
	 	d^{\dagger}_{\vex{q}}
	    c_{\kb}
	 	e^{-i(\kb- \vex{q})\vex{r}_0}
	 	  \right] 
\end{aligned}
\end{align}
describes the tunneling between the metal tip and the superconductor at the contact point $\vex{r}_0$. 
The observable is again the local current at this point [Eq.~(\ref{eq:rfNd})].

Comparing the Hamiltonian of tunneling [Eq.~(\ref{eq:TUNHT})] with that of 
the rf coupling
[Eq.~(\ref{eq:rfHT})]
in the rotating frame $d_{\kb} \rightarrow d_{\kb}e^{-i \omega_L t} $, 
the only difference is that while the contact is global for rf, it is local for tunneling. 
They are identical if the transfer matrix element $\mathcal{T}$ in Eq.~(\ref{eq:TUNHT}) is replaced with 
$\mathcal{T} \delta_{\kb,\vex{q}}$, meaning the momentum is conserved in an rf experiment 
but not in tunneling.

\subsection{General rf and tunneling formulae}

Using linear response theory, we obtain the 
rf current as a function of 
the
coherence factors 
\begin{align}\label{eq:rfI1}
\begin{aligned}
	I_{rf}(t)
	=&2\mathcal{T}^2 \re
	\sum_{\vex{k},\vex{q}} \delta_{\vex{k},\vex{q}}
	\int_{-\infty}^{t}d t'
	e^{i (\omega -\xi_q)(t'-t)}
	\\ &\times
	\left[ 
	(1-f_{\vex{q}}^{(d)})
	v_{\kb}^*(t') v_{\kb}(t)
	-
	f_{\vex{q}}^{(d)}
	u_{\kb}^*(t) u_{\kb}(t')
	\right], 
\end{aligned}
\end{align}
where $\omega \equiv \omega_L - \erf$ is the detuning frequency, and $f_{\kb}^{(d)}$  
is
the
occupation number of atoms 
in state 
$|2\rangle$	
before perturbed by rf radiation. The bare energy 
in the rotating frame is given by
$\xi_k=k^2/2-\masy$. 

Similarly we can obtain the expression for the tunneling current by replacing $\omega$ and 
$\delta_{\vex{k},\vex{q}}$ in Eq.~(\ref{eq:rfI1}) with $-V$ and $1$, respectively. 
In addition, now $f_{\vex{q}}^{(d)} = \theta (q_F -q)$, where $\theta$ denotes the Heaviside 
unit step function. The differential conductance defined as 
$G(t) \equiv -\frac{ d \langle I_{tun} (t) \rangle }{d V}$ is given by
\begin{align}\label{eq:TUNG1}
\begin{aligned}
	G(t) 
    =&\,
    2\mathcal{T}^2 \nu_0 \re
	\sum_{\vex{k}}
    \int_{-\infty}^{t}d t'
    e^{i \tilde{V} (t'-t)}
    \\ &\times
    \left[ 
    	v_{\kb}^*(t') v_{\kb}(t)
    	+
        u_{\kb}^*(t) u_{\kb}(t')
    \right]
	\\
	=&\,
	-
    2\mathcal{T}^2 \nu_0 \im
	\sum_{\vex{k}}
    \int_{-\infty}^{t}d t'
    e^{i \tilde{V} (t'-t)} \,
	\mathcal{G}_{\kb}^{\pup{R}}(t,t'),\!\!
\end{aligned}
\end{align}
where $\tilde{V} \equiv -V-\xi_{q_F}$.
On the last line, we've used the fact that
\[
	\left[ 
    	v_{\kb}^*(t') v_{\kb}(t)
    	+
        u_{\kb}^*(t) u_{\kb}(t')
	\right]
	=
	\langle 
	\!
	\left\{
	\!
	c_{\kb}(t),c_{\kb}^\dagger(t')
	\!
	\right\} 
	\!
	\rangle
	=
	i \mathcal{G}_{\kb}^{\pup{R}}(t,t')
\]
is the single particle retarded Green's function for a BCS product state. 

The rf current [Eq.~(\ref{eq:rfI1})] depends explicitly upon the distribution of the 
$|2\rangle$-state atoms and, as we will see, that of the paired ones as well.
The tunneling conductance [Eq.~(\ref{eq:TUNG1})] by contrast depends
only on the generalized spectral function (Wigner transform of the 
retarded Green's function) in the condensate.

\subsection{RF and tunneling for a Floquet system}

To find the rf signal from a topological Floquet system induced by phase III quench, we 
insert Eq.~(\ref{eq:BdGsol}) into Eq.~(\ref{eq:rfI1}), and obtain the time-averaged rf current
\begin{align}\label{eq:rfIave}
\begin{aligned}
	\bar{I}_{rf} 
	=&\,
	 \pi \mathcal{T}^2
	\sum_{n,\kb}
	\left\lbrace 
	\!\!
	\begin{aligned}
      &\left[ 
       (1-\gamma_{\kb})(1- f_{\kb} ^{(d)})
       -(1+\gamma_{\kb})f_{\kb} ^{(d)}
      \right] 
      \\
	  &\times | \tilde{v}_{n,\kb} |^2
	  \delta{(\omega-\xi_{\kb}-E^{(F)}_{\kb} + n \Omega)}
	  \\
	 +&
      \left[ 
       (1+\gamma_{\kb})(1- f_{\kb} ^{(d)})
       -(1-\gamma_{\kb})f_{\kb} ^{(d)}
      \right] 
      \\
      &\times |\tilde{u}_{n,\kb}|^2
	   \delta{(\omega-\xi_{\kb}+E^{(F)}_{\kb} - n \Omega)}
	   \end{aligned}
		\!
	   \right\rbrace\!,
\end{aligned}
\end{align}
where $\tilde{u}_{n,\kb}$ and $\tilde{v}_{n,\kb}$ are the Fourier coefficients of  
$u_{{\kb}} ^{(F)} (t)$ and $v_{{\kb}} ^{(F)} (t)$:
\begin{align}\label{eq:unvn}
	\begin{bmatrix}
	u_{{\kb}} ^{(F)} (t)
	\\
	v_{{\kb}} ^{(F)} (t)
	\end{bmatrix}
	\equiv \sum_{n=-\infty}^{\infty} 
	\begin{bmatrix}
	\tilde{u}_{n,\kb}
	\\
	\tilde{v}_{n,\kb}
	\end{bmatrix}
	 e^{-i n \Omega t}.
\end{align}
A version of Eq.~(\ref{eq:rfIave}) assuming $f_{\kb} ^{(d)} = 0$
appeared previously in \cite{rfswave}. 
Here all the oscillating terms are discarded since their time averages vanish. 
From an analogous calculation, we determine the 
time-averaged
differential conductance,
\begin{align}\label{eq:TUNGave}
	\bar{G} 
	=
	2 \pi \mathcal{T}^2 \dos
	\sum_{n}
	\sum_{\kb}
	\left[ 
	\begin{aligned}
	| \tilde{v}_{n,\kb} |^2
	\delta{(\tilde{V} -E^{(F)}_{\kb} + n \Omega)}	
	\\
	+
	|\tilde{u}_{n,\kb} |^2
	\delta{ (\tilde{V} + E^{(F)}_{\kb} - n \Omega )}	
	\end{aligned}
	\right].
\end{align}

In a Floquet phase, the rf current and tunneling conductance also show periodic modulation
at harmonics of the
drive
frequency $\Omega$. In Appendix~\ref{Sec:App0}, we 
compute
\begin{align}\label{eq:harmonics}
	I_{rf}(p) \equiv \int_{0}^{T} \frac{d t}{T} e^{i p \Omega t} I_{rf} (t),
	\;
	G(p) \equiv \int_{0}^{T} \frac{d t}{T} e^{i p \Omega t} G (t).
\end{align}
These harmonic amplitudes provide additional observables that can be used to 
characterize a Floquet phase.

The time-averaged rf amplitude in Eq.~(\ref{eq:rfIave}) depends explicitly 
on the distribution function $\gamma_{\kb}$, irrespective of the $|2\rangle$-state occupation $f_{\kb} ^{(d)}$. 
The same conclusion applies for all harmonics [Eq.~(\ref{eq:rfharm})]. By comparison, 
even the instantaneous tunneling conductance $G(t)$ is independent of  $\gamma_{\kb}$, if the
tunneling probe has a constant density of states.

\subsection{ARPES formulae}

In ARPES experiments, one measures the Wigner transform of the single particle lesser Green's function $\G_{\vex{k},<}(t,t')$	
\begin{align}\label{eq:ARPES1}
\begin{aligned}
S(\omega,\kb,t_0)
\equiv 
\int_{-\infty}^{\infty}d t
\,
e^{-i \omega t}
(-i)
\G_{\vex{k},<}\left(t_0-\frac{t}{2},t_0+\frac{t}{2}\right),
\end{aligned}
\end{align}
which equals
\begin{align}\label{eq:Gless}
\begin{aligned}
-i \G_{\vex{k},<}&(t,t')
\equiv	
\langle 
\!
\,
c_{\kb}^\dagger(t')
c_{\kb}(t)
\,
\!
\rangle
=
v_{\vex{k}}^*(t') 
\,
v_{\vex{k}}(t).
\end{aligned}
\end{align}

Inserting Eq.~(\ref{eq:unvn}) into Eq.~(\ref{eq:ARPES1}) and Eq.~(\ref{eq:Gless}), we arrive at the time-averaged ARPES signal from a quench induced Floquet system,
\begin{align}\label{eq:ARPES}
\begin{aligned}
\bar{S}(\omega,\kb)
=&\,
\frac{\pi}{2}
\sum_{n}
\left\lbrace 
\!\!
\begin{aligned}
&
(1-\gamma_{\kb})
| \tilde{v}_{n,\kb} |^2
\delta{(-\omega-E^{(F)}_{\kb} + n \Omega)}
\\
+&
(1+\gamma_{\kb})
|\tilde{u}_{n,\kb}|^2
\delta{(-\omega+E^{(F)}_{\kb} - n \Omega)}
\end{aligned}
\!
\right\rbrace\!.
\end{aligned}
\end{align}

%%%%%%%%%%%%%%%%%%%%%%%%%%%%%%%%%%%%%%%%%%%%%%%%%%%%%%%%%%%%%%%%%%%%%%%%%%%%%%%%%%%%%%%%%%%%%%%%%%%%%%%
%%%%%%%%%%%%%%%%%%%%%%%%%%%%%%%%%%%%%%%%%%%%%%%%%%%%%%%%%%%%%%%%%%%%%%%%%%%%%%%%%%%%%%%%%%%%%%%%%%%%%%%
%%%%%%%%%%%%%%%%%%%%%%%%%%%%%%%%%%%%%%%%%%%%%%%%%%%%%%%%%%%%%%%%%%%%%%%%%%%%%%%%%%%%%%%%%%%%%%%%%%%%%%%
%%%%%%%%%%%%%%%%%%%%%%%%%%%%%%%%%%%%%%%%%%%%%%%%%%%%%%%%%%%%%%%%%%%%%%%%%%%%%%%%%%%%%%%%%%%%%%%%%%%%%%%
%%%%%%%%%%%%%%%%%%%%%%%%%%%%%%%%%%%%%%%%%%%%%%%%%%%%%%%%%%%%%%%%%%%%%%%%%%%%%%%%%%%%%%%%%%%%%%%%%%%%%%%
%%%%%%%%%%%%%%%%%%%%%%%%%%%%%%%%%%%%%%%%%%%%%%%%%%%%%%%%%%%%%%%%%%%%%%%%%%%%%%%%%%%%%%%%%%%%%%%%%%%%%%%
%%%%%%%%%%%%%%%%%%%%%%%%%%%%%%%%%%%%%%%%%%%%%%%%%%%%%%%%%%%%%%%%%%%%%%%%%%%%%%%%%%%%%%%%%%%%%%%%%%%%%%%
%%%%%%%%%%%%%%%%%%%%%%%%%%%%%%%%%%%%%%%%%%%%%%%%%%%%%%%%%%%%%%%%%%%%%%%%%%%%%%%%%%%%%%%%%%%%%%%%%%%%%%%

\section{Weak quenches with constant order parameter: Bulk spectroscopy in phase II \label{Sec:II}}

\subsection{Phase II introduction}

For the long time asymptotic steady state following a phase II quench, only the zeroth order Fourier coefficients 
$\tilde{u}_{0,\kb},\tilde{v}_{0,\kb}$ are nonzero. The time-averaged rf current, tunneling conductance, and ARPES signal can be 
easily determined from Eqs.~(\ref{eq:rfIave}), (\ref{eq:TUNGave}), and (\ref{eq:ARPES}), 
using the exact results for $\Dasy$, $\masy$, and $\gamma(k)$ that can be obtained for a particular quench
\cite{LONG}.
In what follows, unless otherwise stated, the 
non-pairing
state is assumed to be initially unoccupied 
for all momenta in the rf calculation,
$f_{\kb}^{(d)}=0$.

\subsection{Phase II: RF \label{Sec:IIrf}}

\begin{figure}[t]
	\includegraphics[width=0.35\textwidth]{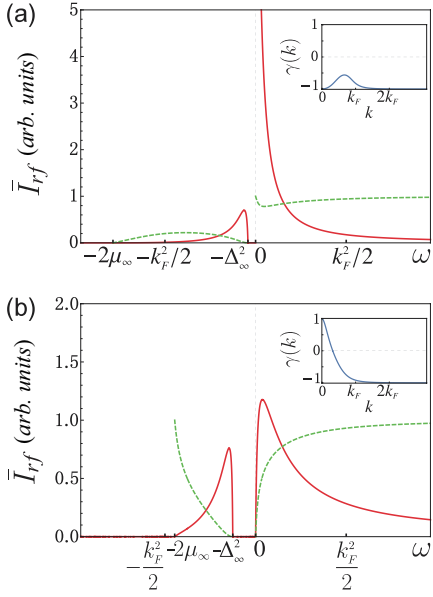}
	\caption{
		(color online).
		Bulk rf spectra (red solid curves) for nonequilibrium states induced by (a) BCS to BCS quench ``a,'' (b) BEC to BCS quench ``b,''
		as indicated in the quench phase diagram Fig.~\ref{fig:diagram}.
		The green dashed curves show the relative weight 
		imparted by the distribution function
		equal to $\frac{1-\gamma(k_\omega)}{2}$, 
		$\frac{1+\gamma(k_\omega)}{2}$ in the positive and negative frequency domain, respectively. 
		The distribution function $\gamma_k$ is illustrated in the inset. 
		It winds from +1 at $k=0$ to -1 as $k \rightarrow \infty$ for 
		the
		BEC to BCS quench b, but goes from -1 to -1 for BCS to BCS quench a. 
		For both quenches, the gap 
		is located
		between $-\Dasy^2$ and $0$ in the spectrum, indicating the 
		(topologically nontrivial) weak-pairing BCS
		nature of 
		the
		post-quench state. A discontinuous jump appears at $\omega=0$ in (a) 
		which is associated to a quench from 
		a
		BCS initial state. 
		The coordinates for quenches a and b are 
		$\lbrace \Di,\Df \rbrace =\lbrace 0.8,0.5\rbrace \Dqcp$ 
		and 
		$\lbrace 1.1,0.8\rbrace \Dqcp $,
		respectively.
	}
	\label{fig:rfII1}
\end{figure}

\begin{figure}[t]
	\includegraphics[width=0.35\textwidth]{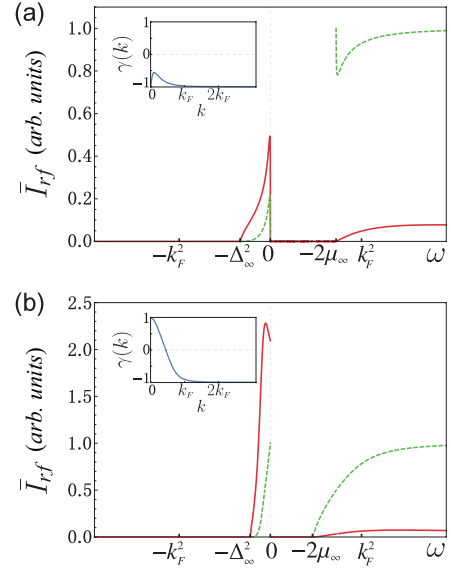}
	\caption{
		(color online).	
		Bulk rf spectra for nonequilibrium states induced by (a) BEC to BEC quench ``c,'' (b) BCS to BEC quench ``d,''
		as indicated in the quench phase diagram Fig.~\ref{fig:diagram}.
		As in
		Fig.~\ref{fig:rfII1}, the red solid and green dashed curves respectively correspond to the actual spectrum and relative weight, 
		while the inset shows the distribution function $\gamma_k$. The distribution function contains 
		an
		even and odd number 
		of zeros for quenches c and d, respectively. The post-quench 
		states corresponding to both (a), (b) are topologically trivial,
		i.e.\ there would be no Majorana edge modes in a system with a boundary,
		and the non-equilibrium phase is ``BEC-like.''
		This is indicated by 
		the gaps 
		which appear
		between $0$ to $|2\masy|$. In (b), the rf spectrum exhibits a discontinuous jump at $\omega=0$,
		indicative of the BCS initial condition.
		The coordinates for quenches c and d are
		$\lbrace \Di,\Df\rbrace =\lbrace 1.02,1.2\rbrace \Dqcp$ 
		and 
		$\left\lbrace 0.7,1.1\right\rbrace \Dqcp$,
		respectively.
	}
	\label{fig:rfII2}
\end{figure}

The rf spectrum of post-quench state with constant order parameter takes 
a
nonzero value only within 
the region
\begin{align}\label{eq:rfgap}
\begin{aligned}
	&
	\omega \in [-2\masy,-\Dasy^2]\cup[0,\infty),\,\,\,
	\masy>\frac{\Dasy^2}{2},
	\\&
	\omega \in [-\Dasy^2,-2\masy]\cup[0,\infty),\,\,\,
	\frac{\Dasy^2}{2}>\masy>0,
	\\&
	\omega \in [-\Dasy^2,0]\cup[|2\masy|,\infty),\,\,\,
	\masy<0.		
\end{aligned}
\end{align}
It is composed of two continuous parts separated by a gap of width $|2\masy|$ or $\Dasy^2$. 
The gap appears for negative (positive) $\omega$ for topologically non-trivial BCS-like (trivial BEC-like) states
with $\masy > 0$ ($\masy < 0$) \cite{LONG,rfpwaveEq}. This feature in principle allows one to distinguish
topological and trivial phases via a bulk measurement, even in the ground state \cite{rfpwaveEq}.

A crucial difference from tunneling is that the weight always extends to zero on one side of the rf gap. 
This is due to the fact that fermions with $k \rightarrow 0$ are not coupled to the order parameter $\Delta$ for 
$p$-wave pairing, and thus transition between hyperfine states at the bare frequency.

The component on the positive (negative) detuning frequency side is due to the process where an rf photon 
with energy $\omega_L$ breaks the ground (excited) Cooper pair labeled by $k_\omega$ and excites one of the 
atoms to the different internal state $\ket{2}$, which does not participate in pairing. An unpaired
state $\ket{1}$ atom is left behind
\cite{LONG}. 
Here
$k_\omega=\sqrt{\frac{\omega(\omega+2\masy)}{\omega+\Dasy^2}}$ is the solution to $\omega = \pm E_k + \xi_k$, 
which comes from the conservation of energy. The term describing this process is weighted by 
$\frac{1-\gamma({k_\omega})}{2}$ $ \left[ \frac{1+\gamma({k_\omega})}{2}\right]$, which reflects the occupation number 
of the ground (excited) Cooper pair.

As required by the conservation of pseudospin winding number 
$Q \equiv s_z(0)-s_z(\infty)$,
the distribution function $\gamma(k)$ must wind from $+1$ at $k=0$ to $-1$ at 
$k \rightarrow \infty$ if the pre- and post-quench states are in different topological phases, i.e.\ 
if 
$\mu_0^{(i)} \masy < 0$ \cite{LONG}. 
When $\mu_0^{(i)}\masy > 0$, $\gamma(k)$ 
does not
wind but approaches $-1$ as 
$k \rightarrow \{0,\infty\}$. 
In particular, the difference between these two 
cases
is the distribution function value at $k=0$ :
\begin{align}\label{|gamma0|}
	\lim_{k \rightarrow 0} 
	\gamma(k)
	=
	\left\{
	\begin{array}{ll}
	-1, & \mui \masy>0, \\
	+1, & \mui \masy<0.
	\end{array}
	\right.
\end{align}

The rf signal from which the distribution function can be extracted therefore possesses topological 
information of the state before and after the quench. A discontinuous jump at $\omega=0$ would appear in 
the rf spectrum if and only if $\mui >0$.  This discontinuous jump exists at the right (left) side of gap 
edge if $\masy>0$ ($\masy<0$). On the other side of the gap, the spectrum grows continuously from 0. 
This introduces the feasibility of using quantum quench in the detection of topological information of the system.

As an example, we evaluated the time-averaged rf spectra of the steady states induced by four different phase 
II quenches. These quenches are indicated as points 
``a'' (BCS to BCS), 
``b'' (BEC to BCS), 
``c'' (BEC to BEC) and 
``d'' (BCS to BEC) 
in Fig.~\ref{fig:diagram}. For quenches a and b (c and d), the asymptotic chemical potential $\masy>\Dasy^2/2$ 
($\masy<0$), and the corresponding rf spectra are illustrated in Fig.~\ref{fig:rfII1} (Fig.~\ref{fig:rfII2}).
The gap 
spans
from $-\Dasy^2$ to $0$ in Fig.~\ref{fig:rfII1}, and 
from
0 to $-2\masy$ in Fig.~\ref{fig:rfII2}, as expected. 
In addition, the discontinuous jump appears at $\omega=0$ 
for quenches a and d, which 
obtain
from 
a
BCS initial state.

\begin{figure}[t]
\includegraphics[width=0.48\textwidth]{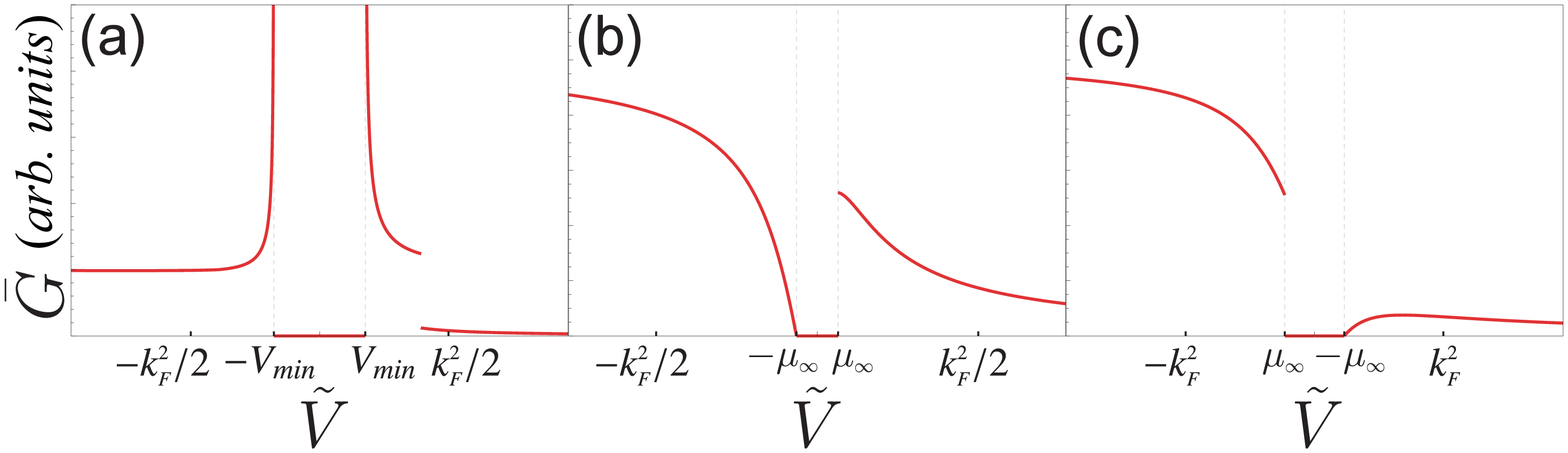}
\caption{The bulk tunneling spectra of the nonequilibrium state following quenches (a) ``a,'' (b) ``f,'' (c) ``d,''
as indicated in the quench phase diagram Fig.~\ref{fig:diagram}.
The corresponding coordinates are 
$\lbrace \Di,\Df\rbrace =\lbrace 0.8, 0.5\rbrace \Dqcp$, 
$\lbrace 0.8,0.9\rbrace \Dqcp$, 
and 
$\lbrace 0.7,1.1\rbrace \Dqcp$, respectively. 
In (a), 
the
long time asymptotic 
nonequilibrium state has
$\masy>\Dasy^2$; two coherence peaks appear at $\tilde{V}=\pm \Vmin$. 
$\Dasy^2>\masy>0$ and $\masy<0$ for 
quenches
f and d respectively. 
In (b), (c), the tunneling signal jumps from 0 at one edge of the gap $\tilde{V}=\masy$ and 
grows continuously at the other edge $\tilde{V}=-\masy$. 
The 
discontinuous 
jump occurs at the right (left) side of the gap in Fig(b) 
[Fig(c)].
}
\label{fig:TUNII}
\end{figure}

\subsection{Phase II: Tunneling \label{Sec:IITun}}

As in rf, a gap appears in the tunneling spectrum. Different from rf, it 
spans the symmetric interval from $-\Emin \leq \tilde{V} \leq \Emin$,
where $\Emin$ is the minimum excitation energy (spectrum gap):
\begin{align}\label{eq:Emin}
	\Emin
	=&
	\left\{
	\begin{array}{ll}
	\Vmin, & \masy>\Dasy^2, \\
	\masy, & \Dasy^2 >\masy>0,\\
	-\masy, & \masy<0,
	\end{array}
	\right.
	\\
	\Vmin \equiv &\,\Dasy \sqrt{2\masy -\Dasy^2}.
\end{align}
Furthermore, the topological properties of
the
post-quench state can also be inferred from tunneling signal near the 
gap edge $\tilde{V}=\pm \Emin$. When $\masy>\Dasy^2$, $\bar{G}(\tilde{V})$ exhibits two coherence peaks around 
$\tilde{V}=\pm \Vmin$, as shown in Fig.~\ref{fig:TUNII} (a). 
The coherence peaks disappear for $\masy<\Dasy^2$, and instead the spectrum shows a discontinuity 
on the positive (negative) edge of the gap for $\masy > 0$ (BCS-like) [$\masy < 0$ (BEC-like)],
see Figs.~\ref{fig:TUNII}(b) and~\ref{fig:TUNII}(c). 
However, unlike the rf, the tunneling spectrum 
does not reveal
information about the pre-quench state 
since the distribution function disappears from the differential conductance.

\subsection{Phase II: ARPES \label{Sec:IIARPES}}

\begin{figure}[t]
	\includegraphics[width=0.5\textwidth]{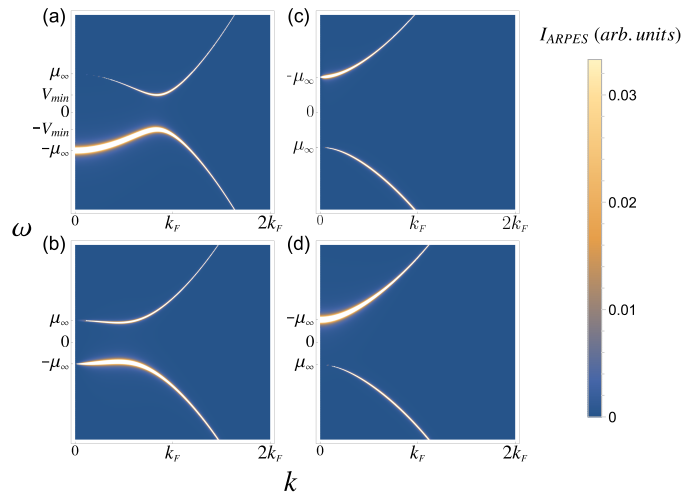}
	\caption{
	(color online).
	The ARPES spectra of the nonequilibrium state induced by quenches (a) ``a,'' (b) ``b,'' (c) ``c,'' (d) ``d.''
	}
	\label{fig:ARPESII}
\end{figure}

There are two branches in the ARPES spectrum of the post-quench state with constant order parameter. 
The upper (lower) one results from the excited (ground) Cooper pair, and is weighted by $\frac{1+\gamma({k_\omega})}{2}$ $ \left[ \frac{1-\gamma({k_\omega})}{2}\right]$. 
For a ground state, 
the upper branch disappears. Similar to tunneling, there is no signal within the region $\omega \in (-\Emin,\,\Emin)$ [see Eq.~(\ref{eq:Emin})]. 
From Eq.~(\ref{eq:ARPES}), the intensity of the signal is determined by both the distribution function $\gamma_{\kb}$ and the moduli
of the coherence factors $|\tilde{u}_{0,\kb}|,|\tilde{v}_{0,\kb}|$. Therefore, the distribution function is also measurable 
from an ARPES experiment.

In Fig.~\ref{fig:ARPESII}, we plot the ARPES signal for the steady state following phase II quenches a, b, c, and d. 
To compute the signal, we replace the delta function in Eq.~(\ref{eq:ARPES}) by a smearing function 
$
	\delta_{sm}(x) \equiv \frac{\eta}{\pi(x^2+\eta^2)}
$, 
and set 
$\eta= (6 \times 10^{-4}) \, \mathcal{E}_F$, where $\mathcal{E}_F = 2 \pi n$ is the Fermi energy of the system
($n$ is the density).
For quenches a and b, the post quench state is topologically nontrivial ($\masy>0$) and, as a result, $|\tilde{u}_{0,\kb}|=0$ at $k=0$. 
Examining the ARPES spectrum at $k=0$, we find it reaches maximum at $\omega=-\masy$ for BCS to BCS quench a 
[Fig~\ref{fig:ARPESII}(a)] and disappears for BEC to BCS quench b [Fig~\ref{fig:ARPESII}(b)]. The absence of signal for 
quench b at $k=0$ is due to the population inversion that occurs when the pre- and post-quench state are in different topological phases.
On the other hand, the post quench states induced by quenches c and d are topologically trivial ($\masy<0$), and 
$|\tilde{v}_{0,\kb}|=0$ at $k=0$. The ARPES signal with $k=0$ maximizes at $\omega=-\masy$ for BCS to BEC quench d 
[Fig~\ref{fig:ARPESII}(d)] and vanishes for BEC to BEC quench c [Fig~\ref{fig:ARPESII}(c)]. For quench d, the peak at 
$\omega=-\masy$ is a signature of the population inversion.

Based on the discussion above, we conclude that topological information of the state before and after the quench 
can be inferred from the ARPES signal. The pre- quench state is in BCS (BEC) phase when the signal is present (absent) 
at $k=0$. In addition, if the peak occurs at negative (positive) frequency, the post-quench state is topologically 
nontrivial (trivial).

%%%%%%%%%%%%%%%%%%%%%%%%%%%%%%%%%%%%%%%%%%%%%%%%%%%%%%%%%%%%%%%%%%%%%%%%%%%%%%%%%%%%%%%%%%%%%%%%%%%%%%%
%%%%%%%%%%%%%%%%%%%%%%%%%%%%%%%%%%%%%%%%%%%%%%%%%%%%%%%%%%%%%%%%%%%%%%%%%%%%%%%%%%%%%%%%%%%%%%%%%%%%%%%
%%%%%%%%%%%%%%%%%%%%%%%%%%%%%%%%%%%%%%%%%%%%%%%%%%%%%%%%%%%%%%%%%%%%%%%%%%%%%%%%%%%%%%%%%%%%%%%%%%%%%%%
%%%%%%%%%%%%%%%%%%%%%%%%%%%%%%%%%%%%%%%%%%%%%%%%%%%%%%%%%%%%%%%%%%%%%%%%%%%%%%%%%%%%%%%%%%%%%%%%%%%%%%%
%%%%%%%%%%%%%%%%%%%%%%%%%%%%%%%%%%%%%%%%%%%%%%%%%%%%%%%%%%%%%%%%%%%%%%%%%%%%%%%%%%%%%%%%%%%%%%%%%%%%%%%
%%%%%%%%%%%%%%%%%%%%%%%%%%%%%%%%%%%%%%%%%%%%%%%%%%%%%%%%%%%%%%%%%%%%%%%%%%%%%%%%%%%%%%%%%%%%%%%%%%%%%%%
%%%%%%%%%%%%%%%%%%%%%%%%%%%%%%%%%%%%%%%%%%%%%%%%%%%%%%%%%%%%%%%%%%%%%%%%%%%%%%%%%%%%%%%%%%%%%%%%%%%%%%%
%%%%%%%%%%%%%%%%%%%%%%%%%%%%%%%%%%%%%%%%%%%%%%%%%%%%%%%%%%%%%%%%%%%%%%%%%%%%%%%%%%%%%%%%%%%%%%%%%%%%%%%

\begin{figure}[b]
\includegraphics[width=0.49\textwidth]{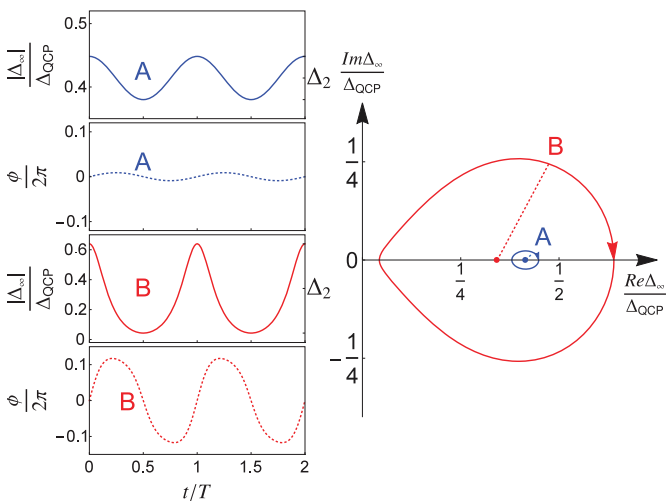}
\caption{
(color online).	
Oscillating order parameter $\Dasy(t)$ induced by two
phase III quenches, labeled ``A'' and ``B''
in the quench phase diagram Fig.~\ref{fig:diagram}.
The left panels show the time dependence of the modulus and phase, while 
the right depicts the $\Dasy(t)$ orbits in the complex $\Delta$ plane.
The magnitude of $|\Dasy(t)|$ oscillates between the two turning points 
$\Delta_2 \pm \Delta_1$ [c.f.\ Eq.~(\ref{eq:Da-mua})].
$\Delta_2$ is indicated by the center point of each orbit on the right. 
The coordinates for quenches A and B are 
$\lbrace \Di,\Df\rbrace =\lbrace 0.119, 0.514\rbrace \Dqcp $, 
and 
$\lbrace 0.00651,0.825\rbrace \Dqcp$, respectively.
}
\label{fig:orbits}
\end{figure}

\section{Strong quenches with periodic order parameter: Bulk spectroscopy in Floquet phase III \label{Sec:III}}

\subsection{Phase III introduction}

The problem of 
computing
the long-time evolution of 
a paired fermion superfluid following a quantum quench
can be reduced to 
solving the integrable dynamics of coupled
Anderson pseudospins. 
Following a similar construction employed in the study of s-wave superfluids 
\cite{YuzbashLax1,YuzbashLax2,YuzbashyanAltshuler06,BarankovLevitov06,DzeroYuzbashyan06},
in \cite{LONG}	
a Lax
spectral method
is used to find the analytical solution
for the 2D p+ip model studied here.
For a system of $N$ pseudospins, one
introduces the spectral polynomial $Q_{2N}(u)$, a conserved integral of motion for any value of $u$. 
In the limit $N \rightarrow \infty$,	
$Q_{2N}(u)$ 
respectively
exhibits zero, one or two isolated pairs of roots in 
phases I, II, III 
of the quench phase diagram, Fig.~\ref{fig:diagram}.
Isolated roots are those separated from the positive real axis in the complex $u$ plane.
These always come in complex conjugate or negative real pairs.
The remaining roots give rise to a branch cut along
the positive real axis in the thermodynamic limit. 

The isolated roots in turn parameterize the dynamics in an
effective ``reduced'' problem of zero, one, or two collective pseudospins.
The asymptotic evolution of the order parameter reaches
a steady state described by Eq.~(\ref{DasyDef}), and this
is completely determined by this reduced solution. 
The dynamics of individual pseudospins can then
be reconstructed following a ``bootstrap'' procedure,
solving the BdG Eq.~(\ref{eq:BdGeq}) and 
exploiting the conservation of $Q_{2N}(u)$ to
determine the occupation of the non-equilibrium spectrum
by ground- or excited-state pairs. 
The explicit solution for phases I and II was provided in 
\cite{LONG}. The analogous calculation for phase III
gives the combination of Floquet states in Eq.~(\ref{eq:BdGsol});
the derivation is relegated to Appendix~\ref{Sec:App1}.

For most quenches in phase III, the spectral polynomial 
possesses
two complex conjugate pairs of isolated roots: one pair 
$u_{1,\pm} \equiv u_{1,\rr} \pm i u_{1,\ii}$ with $u_{1,\rr}>0$ exists only in this phase, 
while the other 
$u_{2,\pm} \equiv u_{2,\rr} \pm i u_{2,\ii}$ 
remains isolated
in both phases II and III. 
Here $u_{a,\rr}$ and $u_{a,\ii}$ indicate the real and imaginary parts of $u_{a,\pm}$, $a=1,2$. 
We define
\begin{align}\label{eq:Da-mua}
	\Delta_a \equiv 
	\sqrt{\frac{|u_a|- u_{a,\rr}}{2}},
	\;\;
	\mu_a \equiv \frac{|u_a|}{2},
	\;\;
	a=1,2,
\end{align}
with $|u_a|\equiv \sqrt{u_{a,\rr}^2+u_{a,\ii}^2}$ 
the modulus of root $u_{a,\pm}$. 
$\Dasy(t) = \Dasy(t + T)$ 
can be expressed in terms of the Jacobi elliptic function 
$\cn(z|M)$
and
the four parameters $\Delta_{1,2}$ and $\mu_{1,2}$. 
The latter are functions of the phase III quench
coordinates $\{\Di,\Df\}$, the fermion density $n$, and
the ultraviolet cutoff energy $\Lambda$.
The period $T$ 
is also determined by the isolated roots \cite{LONG,SHORT}.
In terms of the quench $\{\Di,\Df\}$ 
for $\Di \ll \Df$ and $\Df \lesssim \Dqcp$,
it is given by \cite{SHORT}
\begin{align}\label{Period}
	T 
	\sim 
	{\textstyle{\frac{2}{\Df \sqrt{2 \muf - (\Df)^2}}}} 
	\ln 
	\left[{\textstyle{\frac{8 \pi n}{\Lambda} \frac{\Df \sqrt{2 \muf - (\Df)^2}}{\Di \sqrt{2 \mui - (\Di)^2}}}} \right],
\end{align}
where $\mui$ and $\muf$ denote the ground state chemical potentials corresponding to the 
pre- and post-quench Hamiltonians, respectively.	
In phase II, where $u_{2,\pm}$ 
is the only pair of isolated roots, 
$\masy = \mu_2$ and $\Dasy = \Delta_2$, a constant. 

In the complex $\Delta$ plane, the orbit of $\Dasy (t)$ 
encircles
$\Delta_2$ between the two turning points $\Delta_2+\Delta_1$ and $\Delta_2-\Delta_1$. 
Fig.~\ref{fig:orbits} shows the $\Dasy (t)$ orbits of 
two different phase III quenches ``A'' and ``B,'' 
as indicated in the quench phase diagram Fig.~\ref{fig:diagram}.
The relatively weak quench A is close to phase II, and it 
exhibits
a small orbit, 
with harmonic time-dependence of the amplitude and phase.
The other quench B 
is located
deep in phase III. 
Since the oscillation amplitude is of the same order as the average value,
this constitutes an example of ``strong driving'' with respect to the 
induced Floquet bandstructure. The amplitude and phase of the order parameter
are strongly anharmonic in time for quench B. 
In the remainder of this section, we use these two points as 
representative quenches in phase III.

\subsection{Floquet bandstructure from BCS dynamics}

The explicit solution to the BCS dynamics of the Floquet system induced by 
a
quench in phase III takes the form 
given by
Eq.~(\ref{eq:BdGsol}). It is a superposition of two Floquet states related by 
particle-hole symmetry. We determine the exact expressions 
for these Floquet states and the quasienergy spectrum in Appendix \ref{Sec:App1}.	
We also show that the 
modulus of the
distribution function $|\gamma(k)|$ takes exactly the same form as
previously obtained for phases I and II \cite{LONG}.

\begin{figure}
\includegraphics[width=0.35\textwidth]{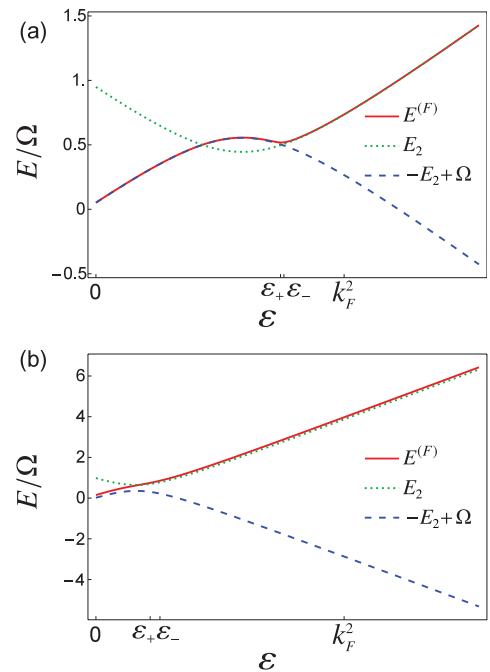}
\caption{
(color online).
Quasienergy spectrum $E^{(F)}(\e)$ in the extended zone scheme for quenches (a) ``A'' and  (b) ``B'',
as indicated in the quench phase diagram Fig.~\ref{fig:diagram}.
Here $\e = k^2$, where $k$ is the momentum.  
For both quenches, the quasienergy  $E^{(F)}(\e)$ (red solid curve) is quite close to 
the static phase II approximation for the quasiparticle excitation energy 
$E_2(\e)$ (green dotted curve) for large $\e$, and to 
$-E_2(\e)+\Omega$ (blue dashed curve) for small $\e$. 
Here $\Omega$ denotes the oscillation frequency of $\Dasy(t)$, determined by the quench.
The crossover of $E^{(F)}(\e)$ between ``ground state'' and ``excited state'' branches
occurs around $\e_\pm$, defined in the text. 
}
\label{fig:QE}
\end{figure}

It will prove very useful to construct ``phase II''
(static quasiequilibrium) analogs of the Floquet states and quasienergies, which will
serve as a reference point for comparison.  
We introduce
\begin{align}\label{eq:IIapproxstates}
	\begin{bmatrix}
	u_2(\e) \\ v_2(\e)
	\end{bmatrix}
	e^{i E_2(\e) t},
	\quad
	\begin{bmatrix}
	v_2 (\e)
	\\ -u_2(\e)
	\end{bmatrix}
	e^{-i E_2(\e) t},
\end{align}
which correspond to the ground and excited Cooper pair states for a phase II quench with 
$\Dasy = \Delta_2$ and $\masy = \mu_2$. 
Here we define	
\begin{align}\label{eq:IIapproxparams}
\begin{aligned}
  	& E_2(\e) \equiv\ \sqrt{(\e/2- \mu_2)^2+\e \Delta_2^2},
  	\\
  	& u_2(\e) \equiv\ \sqrt{\frac{1}{2}+\frac{\e/2 -\mu_2}{2 E_2(\e)}},
  	\\
  	& v_2(\e) \equiv\ -\sqrt{\frac{1}{2}-\frac{\e/2 -\mu_2}{2 E_2(\e)}}.
\end{aligned}
\end{align}
In this section, we denote $\e \equiv k^2$ for modes with momentum 
$\pm \vex{k}$.

\begin{figure}
\includegraphics[width=0.35\textwidth]{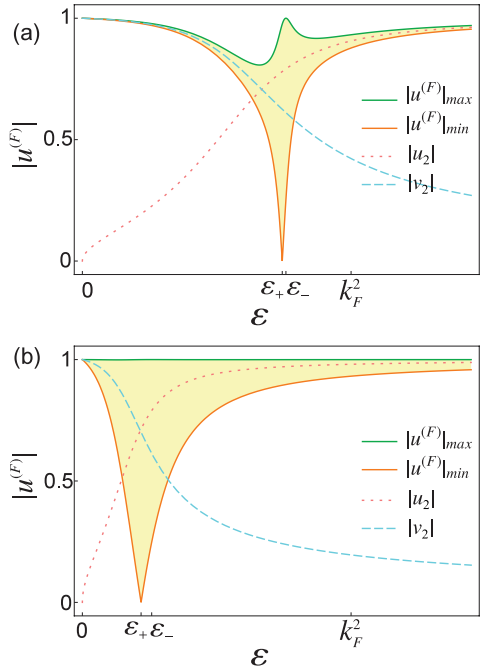}
\caption{
(color online).
Oscillation of the first component of the phase III Floquet solution 
$|u^{(F)}(\e,t)|$, associated to quenches A (a) and B (b).
The top green (bottom orange) solid curve shows the maximum (minimum) of
$|u^{(F)}(\e,t)|$ within one period. The yellow (shaded) area enclosed by these two curves 
marks the region swept out by the periodic modulation. 
We compare these to the static phase II approximation for the coherence factors $|u_2(\e)|$ (pink dotted curve) 
and $|v_2(\e)|$ (cyan dashed curve). 
The magnitude of the Floquet component $|u^{(F)}(\e,t)|$ oscillates within a narrow region around 
the phase II approximation for most single particle energies $\e$, and switches branches near 
$\e_\pm$, where the oscillation amplitude is maximized. 
}
\label{fig:CF}
\end{figure}

In Fig.~\ref{fig:QE}, we plot the quasienergy $E^{(F)}(\e)$ (red solid curve) as a function of 
$\e = k^2$ in the extended zone scheme for quenches A and B. The green dotted curves in these 
figures show the 
dispersion of the
excited states in 
the
phase II approximation $E_2(\e)$, 
while the blue dashed curves are the 
ground state energies shifted by the oscillation frequency 
$-E_2(\e)+\Omega$ [$\Omega = 2\pi /T$, where $T$ is the quench-induced oscillation period of $\Dasy(t)$].
We find that in both cases the quasienergy spectrum exhibits a single ``avoided crossing'' or 
``Floquet bandgap,'' in which the behavior of $E^{(F)}(\e)$ crosses over from the ground to excited
state phase II approximations when the separation between $-E_2(\e)+\Omega$ and $E_2(\e)$ is small
[Fig.~\ref{fig:QE}]. 
In particular, the quasienergy is close to $E_2(\e)$ when $\e \gg \e_-$, and 
to $-E_2(\e)+\Omega$ when $\e \ll \e_+$. 
Here $\e_{\pm}$ ($\e_+ < \e_-$) denote a pair of intermediate single particle energies determined by the quench
[see Eq.~(\ref{eq:epemDEF}) for explicit formulae].  
Technically, $\e_{+}$ ($\e_{-}$) marks the energy at which the coherence factor $u^{(F)}(\e,t)$ [$v^{(F)}(\e,t)$] 
reaches zero within one period [$u^{(F)}(\e_+,0)=0$, $v^{(F)}(\e_-,\frac{T}{2})=0$; 
see Fig.~\ref{fig:CF}, discussed below].

\begin{figure}
\includegraphics[width=0.35\textwidth]{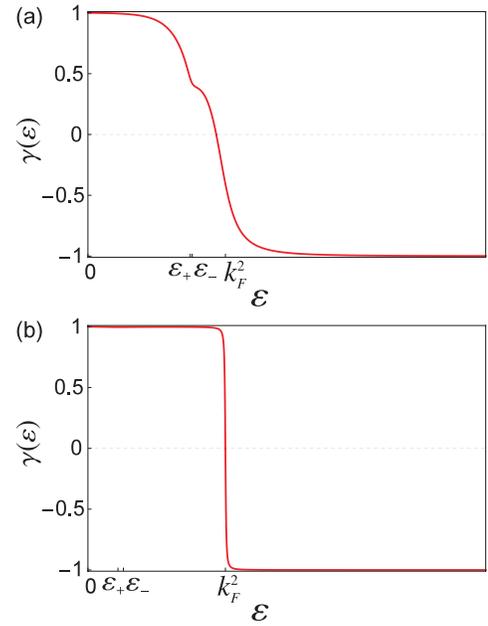}
\caption{
Distribution function $\gamma(\e)$ for quenches (a) A and (b) B. Due to the conservation of 
the pseudospin winding number \cite{LONG}, both wind from $+1$ at $\e=0$ to $-1$ at $\e \rightarrow \infty$.
}
\label{fig:DF}
\end{figure}

The coherence factors of the Floquet state 
$\{
	u^{(F)}(\e,t)
	,\,
	v^{(F)}(\e,t)
\}$ 
oscillate little around those of 
the phase II approximation 
``ground state'' 
$\{
	u_2(\e)
	,\,
	v_2(\e)
\}$ 
or ``excited state'' 
$\{
	v_2(\e)
	,\,
	-u_2(\e)
\}$ 
BdG spinors for most single particle levels, 
switching branches in the same region where the quasienergy changes behavior. 
Fig.~\ref{fig:CF} depicts the oscillation of $|u^{(F)}(\e,t)|$ for quenches A and B. 
The yellow (shaded) area marks the region swept out by $|u^{(F)}(\e,t)|$ in one period. 
It shows that $|u^{(F)}(\e,t)|$ undergoes 
small oscillations near $|v_2(\e)|$ (cyan dashed curve) or $|u_2(\e)|$ (pink dotted curve) 
when $\e \ll \e_+$ or $\e \gg \e_-$, respectively.

As shown in Eq.~(\ref{eq:rfIave}) above, the Floquet bandstructure itself contains 
only part of the information relevant for rf spectroscopy. The other ingredient 
is the non-equilibrium distribution function $\gamma(\e)$, 
as appears in Eq.~(\ref{eq:BdGsol}). This sets the weights of the ``ground'' and ``excited'' 
Floquet state solutions proportional to $\exp\left[\pm i E^{(F)}(\e) t\right]$.
The explicit formula for $\gamma(\e)$ is identical to phases I and II; the latter
was obtained in \cite{LONG}. In Appendix~\ref{Sec:App1}, we confirm that the combination
of Floquet and occupation factors in Eq.~(\ref{eq:BdGsol}) gives the correct
fixed particle density $n$ and self-consistent expression for $\Delta(t)$ in Eq.~(\ref{DasyDef});
see Eq.~(\ref{eq:IIIJE}). 

We plot the distribution functions for quenches A and B in Fig.~\ref{fig:DF}. 
Similar to the quasienergy spectrum and coherence factors, $\gamma(\e)$ 
exhibits a crossover from $-1$ at large $\e$ to $+1$ as $\e \rightarrow 0$. 
This holds true for any phase III quench, and implies that the non-equilibrium
distribution function shows a \emph{population inversion} of the Floquet
states at low energies. I.e., the ``lower'' Floquet band is occupied at large
energies, but the ``upper'' one is filled for $k \lesssim k_F$, 
c.f.\ Eq.~(\ref{eq:fqpDef}).

In fact, the ``winding'' of $\gamma(\e)$ from minus one to plus one as $\e$ decreases
from infinity is required by the topology. In particular, the pseudospin winding number
that characterizes the instantaneous BCS state of the many-fermion system cannot
change following a quench \cite{LONG}. 
(This is different from the question of Majorana edge modes, which can appear or disappear;
these are encoded in the retarded Green's function winding number \cite{LONG,SHORT,Gurarie11},
and this quantity can change in a quench across the quantum critical point \cite{LONG}.)
Given the single crossover of the quasienergy
and Floquet coherence factors relative to the static quasiequilibrium case [phase II approximation,
Eqs.~(\ref{eq:IIapproxstates}) and (\ref{eq:IIapproxparams})], conservation
of the pseudospin winding number for BCS initial states implies that 
$\gamma(\e)$ must go to $+1$ as $\e \rightarrow 0$, so that 
$s^z_{\kb} = \langle c^\dagger_{\kb} c_{\kb} + c^\dagger_{-\kb} c_{-\kb} - 1\rangle/2 = +1/2$
at $\kb = 0$.  

Taking into account both the quasienergy and coherence factors, we find that the Floquet state 
wavefunction in Eq.~(\ref{eq:BdGsol}) 
$\{
	u^{(F)}(\e,t),
	\, 
	v^{(F)}(\e,t)
\}
	e^{i E^{(F)}(\e)t}$ 
can be well-captured by the approximate phase II ground state solution
$\{	
	u_2(\e),
	\, 
	v_2(\e) 
\}
	e^{i E_2(\e) t}$ 
when $\e \gg \e_-$, and the excited state 
$\{ 
	v_2(\e),
	\,
	-u_2(\e)
\}
	e^{-i E_2(\e) t}$ 
when $\e \ll \e_+$. This is true even for the strong quench B. 
As a result, we can construct a full wavefunction for
the phase II approximation as follows:
\begin{align} \label{eq:IIapprox}
\begin{aligned} 
   		\begin{bmatrix}
   		  u^{(II)} (\e,t)
   		  \\
   		  v^{(II)} (\e,t)
   		\end{bmatrix}  
   		& \equiv\,
   	\sqrt{\frac{1-\gamma^{(II)}(\e)}{2}}
   	\begin{bmatrix}
   	 u_2(\e) \\ v_2(\e)
   	\end{bmatrix}
   	e^{i E_2(\e) t}
   	\\
   	&+
   	\sqrt{\frac{1+\gamma^{(II)}(\e)}{2}}
   	\begin{bmatrix}
   	 v_2(\e)
    \\ -u_2(\e)
   	\end{bmatrix}
   	e^{-i E_2(\e) t},
   	\\
  \gamma^{(II)}(\e)& \equiv\sgn(\e-\e_+) \, \gamma(\e), 	
\end{aligned}
\end{align} 
where $\gamma(\e)$ is the true distribution function in phase III. 
The $\sgn$ function is necessary to ``unwind'' the distribution function,
since a phase II quench close to the II--III border has 
$\gamma(\e \rightarrow 0) = -1$ \cite{LONG}.

\subsection{Phase III dynamics, avoided crossing (``Floquet bandgap''), and BCS instability of the normal state \label{Sec:soliton}}

As discussed above, all phase III quenches feature a population imbalance
in the basis of Floquet states at low momenta: the excited state Floquet band
becomes occupied with unit probability in the limit $\e = k^2 \rightarrow 0$. 
This is depicted as the winding from $-1$ to $+1$ with decreasing $\e$ 
of the phase III distribution function $\gamma(\e)$ in Fig.~\ref{fig:DF}.
The mechanism for this is a combination of two factors.
First, the Floquet band structure (quasienergy spectrum and Floquet state
coherence factors) exhibits exactly one ``Floquet bandgap'' or avoided crossing,
as shown in Figs.~\ref{fig:QE} and \ref{fig:CF}. 
In particular, the Floquet coherence factor 
$u^{(F)}(\e,t) \rightarrow 1$ 
$[v^{(F)}(\e,t)] \rightarrow 0$ 
for both $\e \rightarrow \{0,\infty\}$. 
By contrast, $u_2(\e)$ [$v_2(\e)$] approaches one [zero]
for large $\e$, and zero [one] for $\e = 0$; this is the usual
behavior for coherence factors used to parameterize 
positive energy quasiparticle excitations above a 
BCS ground state.  
In the Floquet case, the inversion 
$|u^{(F)}(\e,t)| \rightarrow v_2(\e)$ for $\e \ll \e_+$
implies the winding of $\gamma(\e)$,
so as to ensure the conservation of the
topological pseudospin winding number \cite{LONG}. 

What is the origin of the single avoided crossing in the Floquet bandstructure? 
At first glance, this result is surprising, as we have obtained exact results in a 
system where the (quench-induced) 
drive
frequency $\Omega$ is always much smaller
than the bandwidth (which can be taken as an energy cutoff $\Lambda \gg k_F^2/2$ \cite{LONG}). 
The unperturbed (e.g., phase II approximate) spectrum folds many times
when reduced to the first quasienergy Brillouin zone. We note that the 
periodic drive $\Dasy(t)$ is not in general a pure harmonic
[Fig.~\ref{fig:orbits}]. We might therefore expect the opening of small
bandgaps whenever the folded spectrum approaches the zone edge, in analogy 
with 1D bandstructures in solid state physics. 

In fact, the single crossing can be understood as a consequence of the
integrable BCS dynamics. The key idea is that any quench in phase III
can be ``adiabatically connected'' to a special limiting case. This 
is the limit $\Do^{(i)} \rightarrow 0^+$ for fixed $\Do^{(f)}$,
which describes a quench from the Fermi liquid ground state perturbed
by an infinitesimal seed of $p+ip$ \cite{footnote--p+/-ip} superfluid order. 
In this limit the two isolated pairs 
of roots that characterize all phase III quenches coincide. 
The solution is a single soliton \cite{S-wave1,BarankovLevitov06-II,YuzbashyanAltshuler06} 
in the order parameter. 
For the $p+ip$ case, this takes the form
\begin{align}\label{eq:soliton}
\begin{aligned}
	\Delta(t) =&\, \sqrt{\Rhot(t)} \exp[-i \phi(t)],
	\\
	\Rhot(t)
	=&\,
	\frac{
	2 u_{\ii}^2
	}{
	|u| \cosh\left[2 u_{\ii}\left( t - t_0 \right)\right] + u_{\rr}
	},
\end{aligned}
\end{align}
where $u_{\rr,\ii}$ denote real and imaginary parts of the 
doubly-degenerate isolated roots. 
This soliton solution describes the nonlinear collisionless dynamics
of the Cooper pairs following a linear instability (exponential 
order parameter growth) of the seeded Fermi liquid \cite{S-wave1,BarankovLevitov06-II}.
$\Delta(t)$ grows from the seed at $t \ll t_0$, reaching a maximum
at $t = t_0$; it then decays again as $t \rightarrow \infty$.
The peak magnitude is determined by the post-quench coupling strength;
in terms of $\Df$, one has
\[
	u_{\rr} \simeq 4 \pi n, 
	\quad 
	u_{\ii} \simeq \sqrt{4 \pi n}  \, \Df,
	\quad
	\max \sqrt{\Rhot(t)} \simeq \Df,
\]
valid for $\Df \ll \Dqcp$. Here $n$ denotes the fixed particle density.
The decay $\Delta(t \rightarrow \infty) = 0$
occurs because the quenched system cannot reach the preferred ground state
(a paired BCS superfluid) without dissipating the large energy injected by the quench.
Ultimately, pair-breaking and other processes would induce thermalization, but 
we are neglecting these (see \cite{SHORT} for a discussion of the relevant time scales).

\begin{figure}[t]
\includegraphics[width=0.4\textwidth]{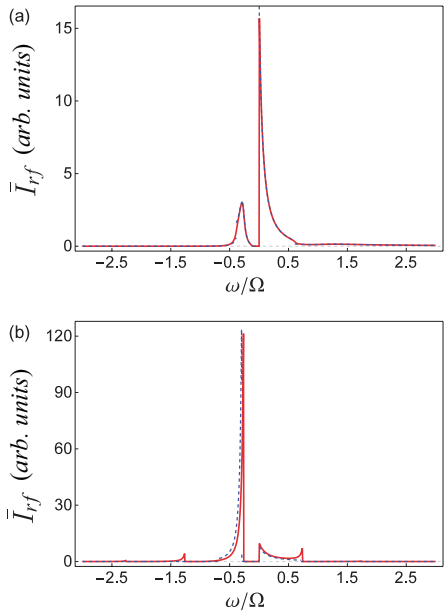}
\caption{
(color online).	
Time-averaged bulk rf spectra (red solid curves) from a topological Floquet system induced by 
the phase III quenches (a) ``A'' and (b) ``B,''
as indicated in the quench phase diagram Fig.~\ref{fig:diagram}.
For comparison, the corresponding phase II (static quasiequilibrium) 
approximations 
[Eq.~(\ref{eq:IIapprox})]
are plotted with dashed blue curves. 
Unsurprisingly, the two curves agree well 
for quench A (a), which is close to the phase II--III border. 
More interesting is the close agreement for quench
B (b), which resides deep in phase III and is characterized
by the large anharmonic oscillations in $\Dasy(t)$ shown in  Fig.~\ref{fig:orbits}.
The good agreement can be largely attributed to the nonequilibrium distribution functions.
These show a population inversion in the Floquet bands at small $\e$, as shown in Fig.~\ref{fig:DF}. 
The small deviation in (b) consists of a sequence of evenly spaced peaks (Floquet copies \cite{rfswave}), 
with the spacing equal to the oscillation frequency $\Omega$. 
}
\label{fig:I0}
\end{figure}

\begin{figure}[t]
	\includegraphics[width=0.35\textwidth]{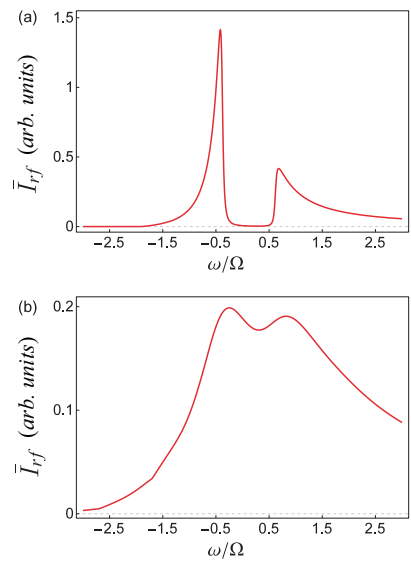}
	\caption{
		Time-averaged bulk rf spectra of the lower Floquet band associated to quenches (a) A and (b) B. 
		These spectra are calculated using the same Floquet states and quasienergies
		utilized
		in Fig.~\ref{fig:I0} but assuming $\gamma(\e)=-1$, that is only the lower Floquet band is occupied. 
		Although these only differ in 
		the
		distribution function, the spectra are completely unlike the 
		corresponding ones in Fig.~\ref{fig:I0}. In particular, the case (b) corresponding to 
		the Floquet bandstructure induced by quench B gives an rf spectrum from the lower Floquet band
		that exhibits no gap, in contrast to quench-induced spectrum in Fig.~\ref{fig:I0}(b),
		which takes into account the physical distribution function.
	}
	\label{fig:NI}
\end{figure}

Mathematically, the soliton solution in Eq.~(\ref{eq:soliton}) can be understood as arising due
to the \emph{discontinuity} in the initial spin distribution of the unperturbed
Fermi step \cite{YuzbashyanAltshuler06}. The only effect of non-zero $\Di$ for
a general phase III quench is to split the pairs of isolated roots, which gives
rise to a train of solitons in time separated by a finite interval $T$,
see Eq.~(\ref{Period}).
By contrast, a normal state initial distribution with multiple discontinuities is expected
to seed a superposition of soliton trains with multiple incommensurate frequencies 
\cite{YuzbashyanAltshuler06}. 

The explicit solution of the Floquet spinors given in Appendix~\ref{Sec:App1}
combined with Eq.~(\ref{eq:soliton}) implies that 
 \[
	|u^{(F)}(\e,t)| \rightarrow 1, 
	\qquad
	|v^{(F)}(\e,t)| \rightarrow 0,
\]
in the limit that $t \rightarrow \infty$, wherein the soliton has decayed. 
One can also verify that $|\gamma(\e)| = 1$ for all $\e$ when
$\Di \rightarrow 0$. Thus, in the long-time limit we recover a pseudospin
configuration consistent with a ``normal state,'' $s^z(\e) = \pm 1/2$. 
This state must possess the same particle density as the initial Fermi step;
moreover, given the connection between discontinuities and isolated roots discussed above,
we expect only a single discontinuity in $s^z(\e)$. We conclude that the 
asymptotic pseudospin texture is that of the initial Fermi step. 
Different from a ground state, however, is the fact that this is encoded
in the sign of the distribution function $\gamma(\e)$, rather than 
the coherence factors (since $|u^{(F)}(\e,t)| = 1$). 
This is again because the quench-induced state is at all times very
far from the preferred ground state. The relation 
$|u^{(F)}(\e,t)| = 1$ for $|t - t_0| \rightarrow \pm \infty$ implies
that the quasiparticle creation operator $\alpha_{\kb}^\dagger \sim c_{\kb}^\dagger$,
where the latter creates an elementary fermion. In other words,
\emph{all} fermions are excitations for a quench that turns on pairing interactions, 
due to the BCS instability for any $k_F > 0$.
These characteristics of the non-equilibrium coherence factors and
distribution function apply to all phase III quenches. 
As can be seen for the deep phase III quench ``B'' shown in 
Fig.~\ref{fig:DF}(b), $\gamma(\e)$ approaches the Fermi step for small $\Di$.

\subsection{Phase III: RF spectroscopy \label{Sec:IIIrf}}

In this subsection, we discuss the rf spectrum of the quench-induced Floquet system and compare it 
with the phase II approximation [Eq.~(\ref{eq:IIapprox})]. Using the exact results for the
Floquet coherence factors and the distribution function, we evaluate Eq.~(\ref{eq:rfIave}).
The time-averaged rf currents 
(red solid curves) for quenches A and B are plotted in Fig.~\ref{fig:I0} together with 
the corresponding 
phase II approximations (blue dashed curves). 
As in Sec.~\ref{Sec:II}, here we assume that
the 
non-pairing
state is initially unoccupied 
for all momenta in the rf calculation,
$f_{\kb}^{(d)}=0$.

\begin{figure}[t]
\includegraphics[width=0.35\textwidth]{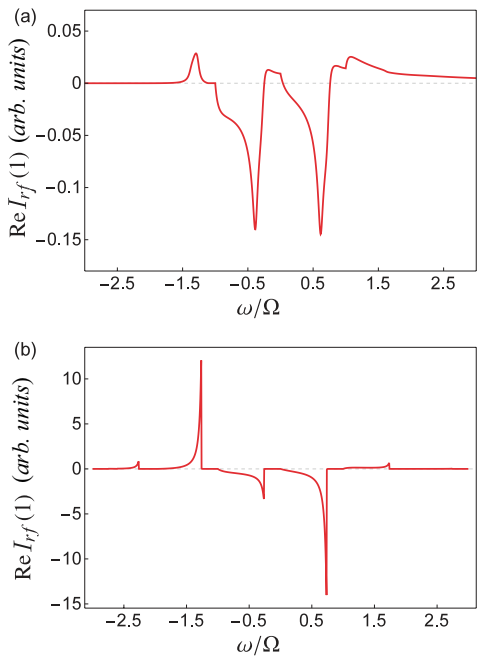}
\caption{
Real part of the bulk first order rf harmonics $\re I_{rf}(1)$ for quenches (a) A, (b) B.
}
\label{fig:I1}
\end{figure}

\begin{figure}
	\includegraphics[width=0.38\textwidth]{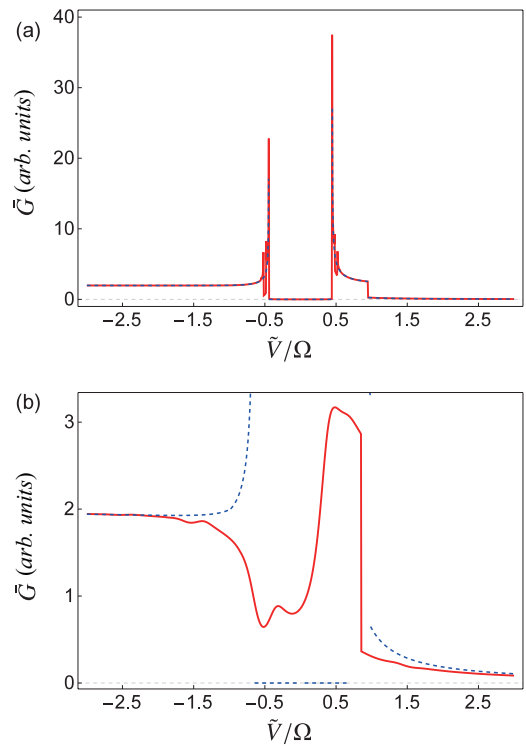}
	\caption{
		(color online).	
		Time-averaged bulk tunneling conductance for a topological Floquet system following quenches 
		(a) A, (b) B. The red solid curve and blue dashed curve in this figures respectively show the 
		phase III conductance $\bar{G}(\tilde{V})$ and its phase II approximation. These two curves agree 
		well in (a) but not in (b). The phase II approximation fails for the strong quench B as the 
		distribution function does not influence the tunneling signal. C.f.\ the rf spectra in 
		Figs.~\ref{fig:I0} and \ref{fig:NI}.
	}
	\label{fig:G0}
\end{figure}

We expect to see 
two
series of peaks evenly spaced by the quench-induced frequency of oscillation $\Omega$. 
These come from processes wherein an rf photon breaks a ground or excited state Cooper pair, and
absorbs or emits several oscillation quanta \cite{rfswave}. No satellite peaks are visible for
quench A, which is quite close to the phase II border and is characterized by a small oscillation amplitude
[see Fig.~\ref{fig:orbits}].
In this case the signal is dominated by the zeroth order term. 
The stronger quench B is deep in phase III; in this case, $\Dasy(t)$ 
exhibits anharmonic time-dependence and a large oscillation amplitude. 
Even for this quench, however, only the lowest order satellite peaks are visible. 
This is because the oscillations in $u^{(F)}(\e,t)$ are confined to a particular region near
$\e_{+} \lesssim \e \lesssim \e_{-}$, as shown in Fig.~\ref{fig:CF}.

In Fig.~\ref{fig:I0}(a), the phase II approximation (blue dashed curve) agrees well with the actual 
rf spectrum for quench A (red solid curve). The deviation is also relatively small for quench B, depicted 
in Fig.~\ref{fig:I0}(b). 
We note in particular that the bulk rf spectrum shows a robust gap, even for this strong 
phase III quench. 
We attribute this to the crucial role played by the distribution function.
As discussed above, the topology of the initial pre-quench state constrains 
$\gamma(\e = 0) = 1$, so that the upper Floquet band is 
occupied at low energies. 
The inversion of $\gamma(\e)$ from $-1$ to $+1$ with decreasing $\e$ effectively interchanges 
$u^{(F)}(\e,t)$ and $v^{(F)}(\e,t)$ in Eq.~(\ref{eq:BdGsol}). 
Since $u^{(F)}(\e,t)$ behaves like $u_2(\e)$ [$v_2(\e)$] for $\e \gg \e_{\pm}$
[$\e \ll \e_{\pm}$], while $v^{(F)}(\e,t)$ shows the opposite behavior, the combination
is well-captured by Eq.~(\ref{eq:IIapprox}).

To compare, we consider a Floquet system where the same oscillating order parameter 
is imposed externally. Unlike the quench-induced case, we assume that this system is 
prepared in a way so that only the lower Floquet band is occupied, i.e.\ $\gamma(\e)=-1$. 
The new time-averaged rf spectrum is shown in Fig.~\ref{fig:NI}, and is found to be 
dramatically different from that of quench induced asymptotic state.  
In particular, for the Floquet states and quasienergy spectrum associated to the strong quench B,
populating the lower Floquet band gives a bulk rf signal that does not exhibit 
a gap, Fig.~\ref{fig:NI}(b). 

In addition to the time-averaged value, the rf signal in the Floquet phase exhibits 
harmonics at the 
drive
frequency. 
In Fig.~\ref{fig:I1}, we plot the real part of the first order bulk rf harmonics signal, 
$\re I_{rf}(1)$. This is defined via Eq.~(\ref{eq:harmonics})
and computed in Eq.~(\ref{eq:rfharm}) in Appendix~\ref{Sec:App0}.
In contrast to the average spectrum, a clear sequences of peaks is observed for
both quenches.

\subsection{Phase III: Tunneling \label{Sec:IIITun}}

Aside from the rf signal, we also compute the time-averaged tunneling conductance, assuming that phase III can 
be realized in a solid. For the quench A located in the vicinity of the phase II--III boundary, the 
phase II approximation works well for most $\tilde{V}$, as shown in Fig.~\ref{fig:G0}(a). 
By contrast, it fails for quench B. In particular, the gap in the static quasiequilibrium phase II approximation 
does not exist in the real spectrum, see Fig.~\ref{fig:G0}(b). 
These results should be compared to the bulk rf spectra, Fig.~\ref{fig:I0} and Fig.~\ref{fig:NI}; the
former uses the physical quench-induced $\gamma(\e)$, while the latter assumes $\gamma(\e) = -1$.
We conclude that the absence of a gap in the tunneling signal for quench 
B can be attributed to the fact that the latter is independent of $\gamma(\e)$, Eq.~(\ref{eq:TUNGave}). 

\begin{figure}[t]
\includegraphics[width=0.3\textwidth]{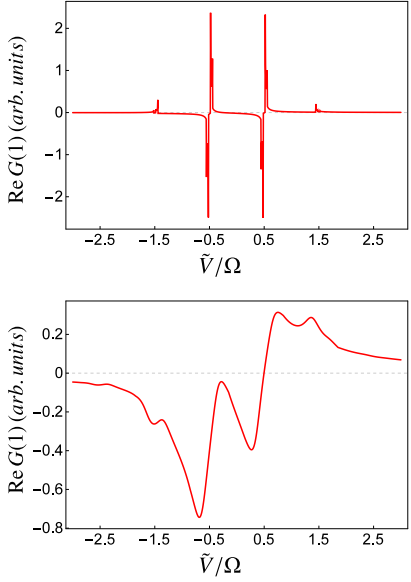}
\caption{
Real part of the bulk first order tunneling harmonics $\re G(1)$ for quenches (a) A, (b) B.
}
\label{fig:G1}
\end{figure}

Fig.~\ref{fig:G1} illustrates the harmonics of the tunneling conductance $\re G(1)$ 
[Eq.~(\ref{eq:harmonics})] for quenches A and B; these are computed via Eq.~(\ref{eq:TUNharm}). 
While clear peaks are observed for the weaker quench A similar to the rf shown in Fig.~\ref{fig:I1}, the
result is much less clear for the strong quench B.

\begin{figure}[b]
\includegraphics[width=0.5\textwidth]{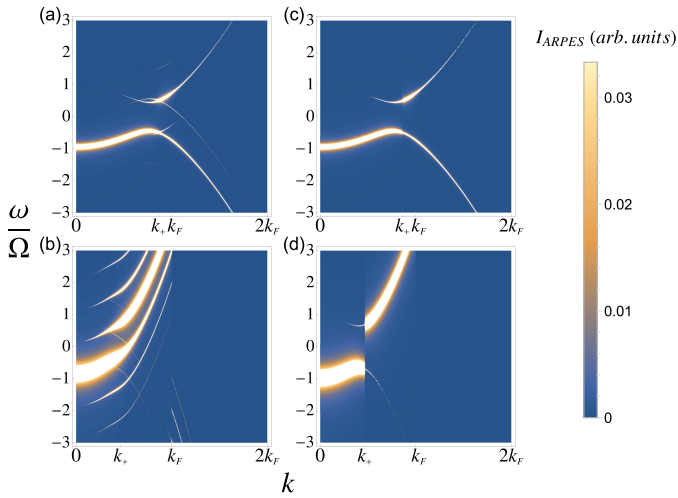}
\caption{
(color online).
The ARPES spectra for the topological Floquet system induced by Phase III quenches (a) A, (b) B. Their phase II approximations are depicted in (c), (d) respectively.
}
\label{fig:ARPESIII}
\end{figure}

\subsection{Phase III: ARPES \label{Sec:IIIARPES}}

The time-averaged ARPES signals from the Floquet system induced by phase III quenches A and B 
are illustrated in Figs.~\ref{fig:ARPESIII} (a) and (b), respectively. 
We find two series of Floquet copies 
in the spectra, and the intensity decreases fast as the Fourier order increases. For comparison, the phase 
II approximations associated with quenches A and B are respectively shown in Fig ~\ref{fig:ARPESIII} (c) 
and (d). We notice the phase II approximation resembles the leading order copy in the ARPES signal for 
most momentum $k$, but not for those around $k_{+}\equiv \sqrt{\e_{+}}$ [Eq.~(\ref{eq:epemDEF})] where 
the wavefunction exhibits the biggest difference from its phase II approximation [Eq.~(\ref{eq:IIapprox})], 
especially for the relatively strong quench B.

\section{Local bulk and edge spectroscopy \label{Sec:Inhomogeneous}}

In this section, we look for experimental signatures of Majorana fermion edge modes 
\cite{Read2000,VolovikBook}
in a 
2D $p + i p$ topological superfluid. 
Specifically, we investigate the contribution from the edge modes to the local rf and tunneling signals. 
Our results apply equally to the BCS ground state, and to a quench-induced topologically non-trivial superfluid in phase II. 
The latter is characterized by a constant $\Dasy$ and $\masy > 0$ [Eq.~(\ref{DasyDef})]; 
quenches producing states of this type are indicated by the shaded portion of region II in Fig.~\ref{fig:diagram}.

The system is assumed to occupy an infinite strip of width $\mathcal{L}$ in the $x$-direction, extended in the $y$-direction.
Our results also apply to the half-plane
$\left\lbrace (x,y)|x \geq 0, y\in \mathbb{R}\right\rbrace $.
At a physical edge $x = x_{\mathsf{Edge}}$, we impose hard wall
boundary conditions.

\subsection{BdG Hamiltonian}

The effective Bogoliubov-de Gennes mean field Hamiltonian in this geometry is given by
\begin{align}\label{eq:rfH0c2}
	H_{\msf{BdG}} =&
	\frac{1}{2}
    \int_{k_y} \int_{x}
    \begin{bmatrix}
     c^{\dagger}_{k_y}(x)
     &
     c_{-k_y}(x)
    \end{bmatrix}
	\hat{h}(k_y)
     \begin{bmatrix}
      c_{k_y}(x)
      \\
      c^{\dagger}_{-k_y}(x)	
     \end{bmatrix}, 
     \nonumber\\
	 \hat{h}(k_y)=&
	 \begin{bmatrix}
	 -\frac{1}{2}\frac{d^2}{d x^2}+\frac{k_y^2}{2} - \masy
	 &
	 \Dasy (-i\frac{d}{dx} - ik_y)
	 \\
	 \Dasy^* (-i\frac{d}{dx} + ik_y)
	 &
	 \frac{1}{2}\frac{d^2}{d x^2}-\frac{k_y^2}{2} + \masy
	 \end{bmatrix}.     
\end{align}
After the Bogoliubov transformation,
\begin{align}\label{eq:atc}
\begin{aligned}
         \begin{bmatrix}
	       a_{k_y} (p) 
	       \\
	       a^{\dagger }_{-k_y} (p)
	     \end{bmatrix}		
	     =
	     \int_{x}
		\begin{bmatrix}
	      u_{k_y} (p,x)
	      &
	      v_{k_y} (p,x)
	      \\
	      v^{*}_{-k_y} (p,x)
	      &
	      u^{*}_{-k_y} (p,x)	    
	     \end{bmatrix}	 
         \begin{bmatrix}
	       c_{k_y} (x) 
	       \\
	       c^{\dagger }_{-k_y}(x) 
	     \end{bmatrix},		    
\end{aligned}
\end{align}
where the amplitudes satisfy
\begin{align}
	\hat{h}^{*}(k_y)
	\begin{bmatrix}
	u_{k_y} (p,x)
	\\
	v_{k_y} (p,x)
	\end{bmatrix}
	=
	E_{k_y}(p)	
	\begin{bmatrix}
	u_{k_y} (p,x)
	\\
	v_{k_y} (p,x)	    
    \end{bmatrix}, 
\end{align}
with
$
	E_{k_y}(p)\geq 0,
$
the Hamiltonian reduces to 
\begin{align}
	H_{\msf{BdG}} 
	=
	&
	\frac{1}{2}
	\int_{k_y} \sum_{p}
	E_{k_y}(p)
	\left[ 
	\begin{aligned}
	&\,
	a^{\dagger}_{k_y}(p) \, a_{k_y} (p)
	\\&\,
	-
	a_{-k_y}(p) \, a^{\dagger }_{-k_y}(p)
	\end{aligned}
	\right]. 
\end{align}
Here $a_{k_y}(p)$ and $a^{\dagger}_{k_y}(p)$ are the annihilation and creation  
operators associated with the quasiparticle excitation labeled by $p$ and 
transverse momentum $k_y$. 
The corresponding quasiparticle energy is denoted as $E_{k_y}(p)$. 
$\sum_{p}$ runs over all the eigenstates with nonnegative eigenvalues,
including edge states (if present).

Atoms in the 
non-pairing
hyperfine state $\ket{2}$ in the rf experiment are 
described by the Hamiltonian
\begin{align}\label{eq:rfH0d2}
\begin{aligned}
	H_{0}^{(d)} =&
    \int_{k_y} \int_{x}
	d_{k_y}^{\dagger}(x) \hat{h}_d(k_y) d_{k_y}(x),\,
	\\
	\hat{h}_d(k_y)=&
	-\frac{1}{2}\frac{d^2}{dx^2} +\frac{k_y^2}{2}- \masy + \erf,
\end{aligned}
\end{align}
which can be reduced to  
\begin{align}
\begin{aligned}
	H_{0}^{(d)}=\int_{k_y}\sum_{k} \left[\xi_{k_y}(k)+\erf \right] b^{\dagger}_{k_y}(k)\, b_{k_y}(k),
\end{aligned}
\end{align}
where $b_{k_y}(k)$ annihilates an atom in state $\ket{2}$,
labeled by $k$ and with energy $\xi_{k_y}(k) + \erf$:
\begin{align}
\begin{aligned}
	& b_{k_y}(k)
	=\,
	\int_{x}
	w(k,x)
	\,
	d_{k_y} (x),
	\\
	& \hat{h}^{*}_d(k_y) \, w(k,x)
	=\left[\xi_{k_y}(k)+\erf\right] w(k,x).
\end{aligned}	
\end{align}
Here $\xi_{k_y}(k) = (k^2 + k_y^2)/2 - \masy$. 
The wavefunction $w(k,x)$ for the 
non-pairing
state is a real 
sinusoid that vanishes at the edges of the cloud. 
In what follows, for clarity we use $p$ and $k$ to indicate the excitation in 
the paired and unpaired atomic levels, respectively.

\subsection{Local rf current}

From the Heisenberg EOM, the local current at $\vex{r}_0$ equals
\begin{align}\label{eq:rfIdef}
\begin{aligned}
	& I(\vex{r}_0,t) 
	\equiv 
	\frac{d n_d(\vex{r}_0,t)}{d t}
	= 
	i \left[
	H,n_d (\vex{r}_0)
	\right],\,
	\\
	& H = H_{\msf{BdG}} + H_{0}^{(d)} + H_T.
\end{aligned}
\end{align}
The local current $I(\vex{r}_0,t)$  therefore is composed of two components from the 
commutator with $H_T$ and $H_{0}^{(d)}$, respectively:
\begin{align}\label{eq:rfIdef2}
\begin{aligned}
	I(\vex{r}_0) 
	=&\, 
	I_T(\vex{r}_0)+I_d(\vex{r}_0),
\\
	I_T(\vex{r}_0)
	\equiv&\,
	i \left[
	H_T, n_d (\vex{r}_0)
	\right],
\\
	I_d(\vex{r}_0)
	\equiv&\,
	i \left[
	H_{0}^{(d)}, n_d (\vex{r}_0)
	\right].
\end{aligned}
\end{align}
In the interaction picture, the current expectation value 
is given by
\begin{align}
	\langle U^{\dagger}(t) \, I(t) \,U(t)\rangle_0
	= 
	I_1
	+
	I_2
	+
	I_3
	+
	\ord{\mathcal{T}^3},
\end{align}
where
\bsub\label{eq:CurComp}
\begin{align}\label{eq:I1}
	 I_1(t)  =& -i \int_{-\infty}^{t} dt'\langle[I_T(t),H_T(t')]\rangle_0
	 ,
	\\
	 I_2(t)  =&\int_{-\infty}^{t} dt_1 \int_{-\infty}^{t} dt_2 \langle H_T(t_1)I_d(t)H_T(t_2) \rangle_0
	  ,
	\\
	\label{eq:I3}
	 I_3(t)  =&-\int_{-\infty}^{t} dt_1 \int_{-\infty}^{t_1} dt_2 
	\left[
		\begin{aligned}
		&\,
		\langle H_T(t_1)H_T(t_2)I_d(t)\rangle_0
		\\&\,
		+ 
		\langle I_d(t)H_T(t_1)H_T(t_2) \rangle_0 
		\end{aligned}
	\right].
\end{align}
\esub
In these equations, 
$\langle ...\rangle_0$ indicates expectation with respect to the state without the rf perturbation. 
In the homogeneous case, only $I_1(t)$ takes a nonzero value. 

We will assume that the initial density of 
non-pairing
state $\ket{2}$ atoms is equal to zero; it is straightforward
to treat the more general case, but the result is rather cumbersome. 
In this case the total current is positive definite, while $I_3(t)$ in Eq.~(\ref{eq:I3}) vanishes exactly. 
For the finite sample geometry of interest here, both $I_{1}(t)$ and $I_2(t)$ are comparable and must be evaluated,
e.g.\ $I_{1}(t)$ can take negative values for a system with an edge.  
We find the following result for the time-averaged local current at $x=x_0$: 
\begin{widetext}
\begin{align}\label{eq:semirf}
\begin{aligned}
	& I(x_0,\omega)
	=
	2 \pi \mathcal{T}^2
    \int_{k_y} \sum_{p,k} 
    \left\lbrace 
	\begin{aligned}
     &
	{\textstyle{\frac{1}{2}}}\left[1 + \gamma_{k_y}(p)\right]
	\left| w(k,x_0)\right| ^2
   	\left| \int_{x} u_{k_y}(p,x)\, w^{*}(k,x)\right| ^2 
	\delta\left[\omega + E_{k_y}(p)-\xi_{k_y}(k)\right]
	 \\
    +
	&
	{\textstyle{\frac{1}{2}}}\left[1 - \gamma_{k_y}(p)\right]
     \left| w(k,x_0)\right| ^2
   	 \left| \int_{x} v_{k_y}(p,x) \, w^{*}(k,x)\right| ^2 
	 \delta\left[\omega - E_{k_y}(p)-\xi_{k_y}(k)\right]
   	 \end{aligned}
   	 \right\rbrace.
\end{aligned} 
\end{align}
\end{widetext}
In this equation, $\gamma_{k_y}(p) = 2 \langle a^{\dagger}_{k_y}(p) \, a_{k_y}(p)\rangle - 1$ encodes
the occupation of the states [induced by the quench; c.f.\ Eq.~(\ref{eq:rfIave})]. 

In the derivation of the local current, we did not use the explicit form of $u_{k_y}(p,x)$, $v_{k_y}(p,x)$, or $w(k,x)$, 
but only the assumption that each is either purely real or imaginary. This assumption 
should be true for any realistic boundary conditions given the form of paired and 
non-pairing
Hamiltonians, 
Eqs~(\ref{eq:rfH0c2}) and (\ref{eq:rfH0d2}). Eq.~(\ref{eq:semirf}) therefore applies to any semi-infinite or infinite system 
governed by Eq.~(\ref{eq:rfH0c2}).

Eq.~(\ref{eq:semirf}) 
takes the form of
Fermi's Golden rule. The first (second) term is due 
to the process where the excited (ground) state 
Cooper pair with 
energy $E_{k_y}(p)$ 
[$-E_{k_y}(p)$]
absorbs an rf photon, 
producing a state $\ket{2}$ atom with energy $\xi_{k_y}(k)$ and an unpaired state $\ket{1}$ atom with 
energy zero. 
The factor 
${\textstyle{\frac{1}{2}}}\left[1 + \gamma_{k_y}(p)\right]$ 
$\left({\textstyle{\frac{1}{2}}}\left[1 - \gamma_{k_y}(p)\right]\right)$ 
encodes 
the probability that the quasiparticle state $\left\lbrace p,k_y\right\rbrace $ is initially occupied (unoccupied) 
before the application of rf radiation. 
The factor $\left| w(k,x_0) \right|^2$ in each term is the probability that an atom exists at $x=x_0$ 
in the non-pairing $\ket{2}$ state with momenta $\left\lbrace k,k_y\right\rbrace $.

Although Eq.~(\ref{eq:semirf}) is very similar to the homogeneous result 
[e.g.\ Eq.~(\ref{eq:rfIave}) 
with only $\tilde{u}_{0,\kb},\tilde{v}_{0,\kb}$ nonzero], it has very different implications for
bulk versus edge states. 
The transition rates are proportional to the squared-overlaps 
\begin{align}
\begin{aligned}
		T_{u;k_y}(p,k) &\equiv \left| \int_{x} u_{k_y}(p,x) \, w^{*}(k,x)\right|^2,
		\\
		T_{v;k_y}(p,k) &\equiv \left| \int_{x} v_{k_y}(p,x) \, w^{*}(k,x)\right|^2.
\end{aligned}
\end{align}
When coherence factors $u_{k_y}(p,x)$ and $v_{k_y}(p,x)$ describe extended bulk states, then the product
of these with extended (box) wavefunction $w^{*}(k,x)$ is of order $1/\mathcal{L}$,
where $\mathcal{L}$ is the extension of the system in the $x$-direction. 
For $p = k$, the integral over $x$ gives a factor of $\mathcal{L}$, so that $T_{u,v;k_y}$ are
of order unity. When $u_{k_y}(p,x)$ and $v_{k_y}(p,x)$ describe a spatially localized edge mode, these are a factor of $1/\mathcal{L}$ smaller.
We conclude that the contribution of the edge states is suppressed by a factor of the 
linear system size. Moreover, because $k_y$ runs over a finite range for the edge modes,
the edge signal is delocalized throughout the bulk of the sample (since it obtains from 
integrating $|w(k,x_0)|^2$ over a finite range of $k$ for which energy conservation is satisfied). 

Thus, local rf spectroscopy of this type is not well-suited to detect Majorana edge modes.

\subsection{Tunneling}

By contrast, the local tunneling conductance is given by
\begin{multline}\label{eq:semiTUN}
	G(x_0,\tilde{V})
	\\
	=
	2 \pi \mathcal{T}^2 \nu_0
	\int_{k_y}
	\sum_{p}
	\left\{
	\begin{aligned}
	&\,
	\left| u_{k_y}(p,x_0)\right| ^2
	\delta{\left[\tilde{V}+E_{k_y}(p)\right]}
		\\
		&
	+
	\left| v_{k_y}(p,x_0)\right| ^2
	\delta{\left[\tilde{V}-E_{k_y}(p)\right]}
	\end{aligned}
	\right\}\!.
\end{multline}
This is independent of the distribution function (quench-induced occupancy), and
gives equal weight to discrete bound states (edge modes)
and continuum bulk modes. The contribution of each of the latter is suppressed by a factor
of $1/\mathcal{L}$, but this is compensated by the summation over $p$. 

When the metal tip is placed deep in the bulk, the tunneling spectrum resembles that of the 
homogeneous case. Near the boundary, signal within the gap is contributed only by the edge modes.

%%%%%%%%%%%%%%%%%%%%%%%%%%%%%%%%%%%%%%%%%%%%%%%%%%%%%%%%%%%%%%%%%%%%%%%%%%%%%%%%%%%%%%%%%%%%%%%%%%%%%%%
%%%%%%%%%%%%%%%%%%%%%%%%%%%%%%%%%%%%%%%%%%%%%%%%%%%%%%%%%%%%%%%%%%%%%%%%%%%%%%%%%%%%%%%%%%%%%%%%%%%%%%%
%%%%%%%%%%%%%%%%%%%%%%%%%%%%%%%%%%%%%%%%%%%%%%%%%%%%%%%%%%%%%%%%%%%%%%%%%%%%%%%%%%%%%%%%%%%%%%%%%%%%%%%
%%%%%%%%%%%%%%%%%%%%%%%%%%%%%%%%%%%%%%%%%%%%%%%%%%%%%%%%%%%%%%%%%%%%%%%%%%%%%%%%%%%%%%%%%%%%%%%%%%%%%%%
%%%%%%%%%%%%%%%%%%%%%%%%%%%%%%%%%%%%%%%%%%%%%%%%%%%%%%%%%%%%%%%%%%%%%%%%%%%%%%%%%%%%%%%%%%%%%%%%%%%%%%%
%%%%%%%%%%%%%%%%%%%%%%%%%%%%%%%%%%%%%%%%%%%%%%%%%%%%%%%%%%%%%%%%%%%%%%%%%%%%%%%%%%%%%%%%%%%%%%%%%%%%%%%
%%%%%%%%%%%%%%%%%%%%%%%%%%%%%%%%%%%%%%%%%%%%%%%%%%%%%%%%%%%%%%%%%%%%%%%%%%%%%%%%%%%%%%%%%%%%%%%%%%%%%%%
%%%%%%%%%%%%%%%%%%%%%%%%%%%%%%%%%%%%%%%%%%%%%%%%%%%%%%%%%%%%%%%%%%%%%%%%%%%%%%%%%%%%%%%%%%%%%%%%%%%%%%%

\section{Conclusion \label{Sec: End}}

In this paper, we have obtained the rf and tunneling spectra of a quench induced out-of-equilibrium steady state. 
In particular, we focused on the Floquet system with time periodic order parameter induced by a phase III quench. 
One important aspect of our result is the connection between these experimental observables and the distribution function. 
We found that the distribution function plays an essential role in the bulk rf spectrum. 
We demonstrated that the reason the time averaged rf (but not tunneling) spectra are in good agreement with 
static quasiequilibrium (phase II) approximations lies in the distribution function, which is 
forced to exhibit a population inversion in the basis of Floquet states due to the conservation of the 
topological pseudospin winding number. This crucial information is missing in 
the 
tunneling spectrum, and leads
to the disappearance of the gap in tunneling spectra for strong phase III quenches characterized 
by large amplitude oscillations in the order parameter.  
Finally, we showed that local rf is not a good method to detect Majorana edge states, due to the 
non-local character of the radiation-induced transitions to non-pairing states.

\begin{acknowledgments}

We thank Luca D'Alessio, Victor Gurarie, Kaden Hazzard, and Emil Yuzbashyan for useful discussions. 
We acknowledge funding from the Welch Foundation under Grant No. C-1809 and from an Alfred P. Sloan Research Fellowship (No. BR2014-035).

\end{acknowledgments}

\appendix

\section{RF and tunneling amplitude harmonics \label{Sec:App0}}

In the superfluid Floquet phase, 
the rf current given by Eq.~(\ref{eq:rfI1}) exhibits modulations
at harmonics of the 
drive 
frequency $\Omega$. These are 
encoded in
\begin{widetext}
\begin{align}\label{eq:rfharm}
	I_{rf}(p) 
	=&\,
	\frac{ \pi \mathcal{T}^2}{2}
	\sum_{n,\kb}
	\left\lbrace 
	\begin{aligned}
      &\left[ 
       (1-\gamma_{\kb})(1- f_{\kb} ^{(d)})
       -(1+\gamma_{\kb})f_{\kb} ^{(d)}
      \right] 
       ( \tilde{v}_{n,\kb} ^ * \tilde{v}_{n+p,\kb} 
      	  + \tilde{v}_{n-p,\kb} ^ *
      	\tilde{v}_{n,\kb})   
	 \delta{(\omega-\xi_{\kb}-E^{(F)}_{\kb} + n \Omega)}     
      \\
      &
	+
      \left[ 
       (1+\gamma_{\kb})(1- f_{\kb} ^{(d)})
       -(1-\gamma_{\kb})f_{\kb} ^{(d)}
      \right] 
		( \tilde{u}_{n,\kb} ^ * \tilde{u}_{n+p,\kb}
		+ \tilde{u}_{n-p,\kb} ^ * \tilde{u}_{n,\kb})
	 \delta{(\omega-\xi_{\kb}+E^{(F)}_{\kb}-n \Omega)}	      
	\end{aligned}
	\right\rbrace
	\nonumber\\
	-&
	\frac{ i \mathcal{T}^2}{2}
	\sum_{n,\kb}
	\left\lbrace 
	\begin{aligned}
      &\left[ 
       (1-\gamma_{\kb})(1- f_{\kb} ^{(d)})
       -(1+\gamma_{\kb})f_{\kb} ^{(d)}
      \right] 
       ( \tilde{v}_{n,\kb} ^ * \tilde{v}_{n+p,\kb} 
      	  - \tilde{v}_{n-p,\kb} ^ *
      	\tilde{v}_{n,\kb})    
	\frac{1}{\omega-\xi_{\kb}-E^{(F)}_{\kb} + n \Omega}
      \\
      &-
      \left[ 
       (1+\gamma_{\kb})(1- f_{\kb} ^{(d)})
       -(1-\gamma_{\kb})f_{\kb} ^{(d)}
      \right] 
		( \tilde{u}_{n,\kb} ^ * \tilde{u}_{n+p,\kb}
		- \tilde{u}_{n-p,\kb} ^ * \tilde{u}_{n,\kb})
		 \frac{1}{\omega-\xi_{\kb}+E^{(F)}_{\kb}-n \Omega}      
	\end{aligned}
	\right\rbrace.
\end{align}
An analogous expression for harmonics of the 
tunneling conductance is given by
\begin{align}\label{eq:TUNharm}
\begin{aligned}
	G(p) 
	=&\,
	 \pi \mathcal{T}^2 \dos
	\sum_{n,\kb}
\left[
	( \tilde{v}_{n,\kb} ^ * \tilde{v}_{n+p,\kb} 
		+ \tilde{v}_{n-p,\kb} ^ * \tilde{v}_{n,\kb})
	 \delta{(\tilde{V} - E^{(F)}_{\kb} + n \Omega)}
+
	( \tilde{u}_{n,\kb} ^ * \tilde{u}_{n+p,\kb}
	+ \tilde{u}_{n-p,\kb} ^ * \tilde{u}_{n,\kb})
	\delta{ (\tilde{V} + E^{(F)}_{\kb} - n \Omega)} 
\right]
	\\
	&-
	i \mathcal{T}^2 \dos
	\sum_{n,\kb}
\left[ 
	( \tilde{v}_{n,\kb} ^ * \tilde{v}_{n+p,\kb} 
		- \tilde{v}_{n-p,\kb} ^ * \tilde{v}_{n,\kb})
	\frac{1}{\tilde{V} - E^{(F)}_{\kb} + n \Omega}
-
	( \tilde{u}_{n,\kb} ^ * \tilde{u}_{n+p,\kb}
	- \tilde{u}_{n-p,\kb} ^ * \tilde{u}_{n,\kb})
	\frac{1}{\tilde{V} + E^{(F)}_{\kb} - n \Omega}	
\right]. 
\end{aligned}
\end{align}
\end{widetext}
We find that $\tilde{u}_{n,\kb},\tilde{v}_{n,\kb}$ are real, so the second term in 
Eqs.~(\ref{eq:rfharm}) and (\ref{eq:TUNharm}) can be ignored if only the real parts of 
$I_{rf}(p)$ and $G(p)$ are needed.

%%%%%%%%%%%%%%%%%%%%%%%%%%%%%%%%%%%%%%%%%%%%%%%%%%%%%%%%%%%%%%%%%%%%%%%%%%%%%%%%%%%%%%%%%%%%%%%%%%%%%%%
%%%%%%%%%%%%%%%%%%%%%%%%%%%%%%%%%%%%%%%%%%%%%%%%%%%%%%%%%%%%%%%%%%%%%%%%%%%%%%%%%%%%%%%%%%%%%%%%%%%%%%%
%%%%%%%%%%%%%%%%%%%%%%%%%%%%%%%%%%%%%%%%%%%%%%%%%%%%%%%%%%%%%%%%%%%%%%%%%%%%%%%%%%%%%%%%%%%%%%%%%%%%%%%
%%%%%%%%%%%%%%%%%%%%%%%%%%%%%%%%%%%%%%%%%%%%%%%%%%%%%%%%%%%%%%%%%%%%%%%%%%%%%%%%%%%%%%%%%%%%%%%%%%%%%%%
%%%%%%%%%%%%%%%%%%%%%%%%%%%%%%%%%%%%%%%%%%%%%%%%%%%%%%%%%%%%%%%%%%%%%%%%%%%%%%%%%%%%%%%%%%%%%%%%%%%%%%%
%%%%%%%%%%%%%%%%%%%%%%%%%%%%%%%%%%%%%%%%%%%%%%%%%%%%%%%%%%%%%%%%%%%%%%%%%%%%%%%%%%%%%%%%%%%%%%%%%%%%%%%
%%%%%%%%%%%%%%%%%%%%%%%%%%%%%%%%%%%%%%%%%%%%%%%%%%%%%%%%%%%%%%%%%%%%%%%%%%%%%%%%%%%%%%%%%%%%%%%%%%%%%%%
%%%%%%%%%%%%%%%%%%%%%%%%%%%%%%%%%%%%%%%%%%%%%%%%%%%%%%%%%%%%%%%%%%%%%%%%%%%%%%%%%%%%%%%%%%%%%%%%%%%%%%%
%%%%%%%%%%%%%%%%%%%%%%%%%%%%%%%%%%%%%%%%%%%%%%%%%%%%%%%%%%%%%%%%%%%%%%%%%%%%%%%%%%%%%%%%%%%%%%%%%%%%%%%
%%%%%%%%%%%%%%%%%%%%%%%%%%%%%%%%%%%%%%%%%%%%%%%%%%%%%%%%%%%%%%%%%%%%%%%%%%%%%%%%%%%%%%%%%%%%%%%%%%%%%%%
%%%%%%%%%%%%%%%%%%%%%%%%%%%%%%%%%%%%%%%%%%%%%%%%%%%%%%%%%%%%%%%%%%%%%%%%%%%%%%%%%%%%%%%%%%%%%%%%%%%%%%%
%%%%%%%%%%%%%%%%%%%%%%%%%%%%%%%%%%%%%%%%%%%%%%%%%%%%%%%%%%%%%%%%%%%%%%%%%%%%%%%%%%%%%%%%%%%%%%%%%%%%%%%

\section{Floquet states and occupations via integrability: Explicit solution in Phase III \label{Sec:App1}}

Through self-consistent mean field theory and the use of the Lax construction, 
the order parameter $\Delta(t)$ as a function of time is determined
($ \Delta \equiv  \sqrt{\Rhot} e^{-i \phi} $).
The amplitude ($\sqrt{ \Rhot}$) and the argument ($-\phi$) 
of the complex order parameter follow the EOM \cite{LONG}
\begin{align}\label{eq:RhotEOM}
	\dot{\Rhot}^2
	=
	(\Rhot_{+} - \Rhot)(\Rhot - \Rhot_{-})
	(\Rhot + \wRhot_{+})
	(\Rhot + \wRhot_{-}),
\end{align}

\begin{align}\label{eq:PhiEOM}
	\dot{\phi} 
	=
	\frac{3}{2}
	\Rhot
	+
	2 \muco
	-	
	\frac{\psi}{\Rhot}.
\end{align}
where
\begin{align}\label{eq:RhotParams}
\begin{aligned}
	\Rhot_{\pm}
	\equiv&\,
	\left(
	  \Delta_1 \pm \Delta_2
	\right)^2,
	\\
	\wRhot_{\pm} 
	\equiv&\,
	\left(
	\sqrt{2 \mu_1 -\Delta_1^2} \pm \sqrt{2 \mu_2 -\Delta_2^2}	
	\right)^2,
\end{aligned}
\end{align}
\begin{align}\label{eq:PhiParams}
\begin{aligned}
	\muco 
	\equiv&\,
	\frac{1}{2}
	\left(
	 \mu_1+\mu_2 -\Delta_1^2-\Delta_2^2
	\right),
	\\
	\psi \equiv&\,
		\frac{1}{2}
	\left( \Delta_1^2 -\Delta_2^2	\right)
	\left( \Delta_1^2 -\Delta_2^2 -2 \mu_1 +2\mu_2
		\right).
\end{aligned}
\end{align}
The parameters $\Delta_{1,2}$ and $\mu_{1,2}$ are determined
via Eq.~(\ref{eq:Da-mua}) 
by the two pairs of isolated roots that characterize a phase III quench.
For a given quench specified by the coordinates
$\{\Di,\Df\}$, these roots solve a certain 
transcendental equation \cite{LONG}. 
Results can be obtained numerically for a given fixed particle density $n$ 
and ultraviolet energy cutoff $\Lambda$.

The solution $\phi (t)$ to Eq.~(\ref{eq:PhiEOM}) is a combination of a 
periodic part $\Phi (t)$ and linear part $2 \masy t$. 
Here $\Phi (t)$ shares the same period $T$ with $\Rhot (t)$, and $\masy$ is defined as
\begin{align}\label{eq:masy}
	\masy
	\equiv
	\frac{1}{2 T}
	\int_0^T d t 
	\, \dot{\phi}
	=	
	\frac{1}{2 T}
	\int_0^T d t 
	\,
	\left( 
	\frac{3}{2}\Rhot(t)
	+
	2 \muco
	- \frac{\psi}{\Rhot (t)}
	\right). 
\end{align}

\subsubsection{Lax reduced solution}

In the following, a ``Lax reduced'' solution \cite{LONG} to the BCS spin dynamics 
is utilized to find the explicit solution to our problem, 
i.e.\ the asymptotic steady state following a sudden quench of coupling strength.

We consider a reduced solution with the same order parameter (isolated roots) as a particular phase III quench, 
but with different initial conditions. 
The existence of such solutions, along with an explicit prescription for constructing them,
was detailed in Secs.~III C and IV B of \cite{LONG}, based upon previous s-wave work \cite{YuzbashLax2}.
In terms of Anderson pseudospins, the reduced solution can be written as a product of a sign function $\zeta(\e)$ 
and a pseudospin function $\vec{s}_{0}(\e)$,
\begin{align}\label{eq:ResTor0}
	    \vec{s}_{R}(\e)
	    =&\,
	    \zeta(\e)
	    \vec{s}_{0}(\e),
\end{align}
where $\zeta(\e) = \pm 1$. 
The single particle levels are labeled via the momentum-squared $\e \equiv k^2$. 
To qualify as a solution, $\zeta(\e)$ will exhibit 
a discontinuous jump at some $\e$ from $+1$ to $-1$; this is closely related
to the soliton solution discussed above in Sec.~\ref{Sec:soliton}. 
The function
 $\vec{s}_{0}(\e)$ depends only on the order parameter, and is given by
\begin{align}\label{eq:ResTor}
\begin{aligned}
	s_{0}^z (\e)
	=&
	a(\e) \Rhot
	+
	b(\e),
	\\
    s_{0}^- (\e) 
	=&
	\frac{a(\e)}{2\sqrt{\e \Rhot}}
    \left[ 
	 - i \dot{\Rhot}
	 +
     \Rhot^{2}+ (4 \muco - 2 \e) \Rhot +2 \psi
    \right] 
  	e^{-i \phi},
\end{aligned}
\end{align}
where $a(\e)$ and $b(\e)$ are functions of $\e$,
\begin{align}\label{eq:abDef}
\begin{aligned}
	a(\e)
	=&\,
	\frac{ \e }
	{4 E_1(\e) E_2(\e)},
	\\
	b(\e)
	=&\,
	-\frac{E_1^2(\e)+E_2^2(\e) -(\mu_1 - \mu_2)^2}
	{4 E_1(\e) E_2(\e)}.
\end{aligned}
\end{align}
Here $E_a(\e) \equiv \sqrt{(\e/2 - \mu_a)^2 + \Delta_a^2 \e}$, $a=1,2$
is a $p+ip$ quasiparticle energy [Eq.~(\ref{Ek})]. 

The order parameter is defined self consistently by 
\begin{align}\label{eq:Delta}
	\Delta(t) \equiv-G \sum_i \sqrt{\e_i} s_i^-.
\end{align}
Substituting $s_i^-$ in this equation from the reduced solution 
in Eq.~(\ref{eq:ResTor0}) and using Eq.~(\ref{eq:ResTor}), we find that the 
L.H.S.\ reduces to 
\begin{align}
\begin{aligned}
	&-G(\sum_i \zeta_i a_i)
	 \frac{ - i \dot{\Rhot}
		 +
	     \Rhot^{2}+ 4 \muco \Rhot +2 \psi
	    }{2 \sqrt{\Rhot}}
	  	e^{-i \phi}
	  	\\
	 & +G
    (\sum_i \zeta_i \e_i a_i)\sqrt{\Rhot}
    e^{-i \phi}.	  	
\end{aligned}	
\end{align}
This gives $\sqrt{\Rhot}e^{-i\phi}$ if and only if
\begin{align}\label{eq:ResTorCons}
\begin{aligned}
	 \sum_i \zeta_i a_i=\,0 
	,\quad
	\sum_i \zeta_i \varepsilon_i a_i=\,\frac{1}{G},
\end{aligned}
\end{align}
which serve as the constraints for the Lax reduced solution.
Under such constraints, we find the total z-spin $J$ 
(related to the fixed number of particles $N$) 
and the total energy $E$ are indeed conserved quantities since
\begin{align}\label{eq:ResTorJE}
\begin{aligned}
	J \equiv &\,
	\sum_{i} 
	\zeta_i 
	s_{0,i}^z
	=\,
	\sum_{i} 
	\zeta_i b_{i}, 
	\\
	E \equiv &\,
	\sum_{i} 
	\zeta_i 
	\varepsilon_i
	s_{0,i}^z
	-\frac{\Rhot}{G}
	=\,
	\sum_{i} 
	\zeta_i
	\varepsilon_i  b_{i}.
\end{aligned}
\end{align}

Another important quantity is the Lax norm \cite{LONG}, which generates all the integrals of motion. 
We find that the Lax norm of the reduced solution is given by
\begin{align}\label{eq:ResTorL2}
	L_2 (u) =\,
	\left( 
	  \sum_i 
	   \frac{\zeta_i \e_i }
	     {2 \sqrt{ \mathcal{Q}_4(\e_i) }}
	     \frac{1}{\e_i - u}
	     \right) ^2
	     \mathcal{Q}_4(u).	
\end{align}
The $\left\lbrace \zeta_i\right\rbrace $ can in principle be determined by equating 
Eq.~(\ref{eq:ResTorL2}) to its initial value.

To find the spinor of $\vec{s}_0$ parameterized as 
\begin{align}
	\begin{bmatrix}
	  u_{0} (\e,t)
	  \\
  	  v_{0} (\e,t)
	\end{bmatrix}
	= &\,
	\begin{bmatrix}
	 | u_{0}(\e,t) | 
	 e^{- i \theta_{u}(\e,t)}
	 \\
	 | v_{0}(\e,t) |
	 e^{- i \theta_{v}(\e,t)}
	\end{bmatrix},
\end{align}
we use
\begin{align}\label{eq:spin2coh}
\begin{aligned}
	s_0^-(\e,t) 
	=&\,
	u_0^*(\e,t) 
	v_0(\e,t) 
	\\
	s_0^z (\e,t) 
	=&\,
	\frac{1}{2}
	\left( 
	 | v_0(\e,t)  |^2
	 -| u_0(\e,t)  |^2
	 \right).
\end{aligned}	 
\end{align}
In the rotating frame 
$s^{-} \rightarrow s^{-}e^{i \phi }$, 
this yields
\begin{align}\label{eq:absuv}
\begin{aligned}
	| u_{0} (\e,t) |
	=&\,
	\sqrt{\frac{1}{2} -
	\left[
	a(\e) \Rhot (t)
	+ b(\e)
	\right] },
	\\
	| v_{0}(\e,t)|
	=&\,
	\sqrt{\frac{1}{2} +
	\left[ 
	a(\e) \Rhot (t)
	+ b(\e)
	\right] },
\end{aligned}
\end{align}
\begin{align}\label{eq:thetauv}
	\theta_{u}(\e,t)-\theta_{v}(\e,t)
	=
	\arg 
	\left( 
	 -i \dot{\Rhot}
	 +\Rhot^2 + 4 \muco  \Rhot
	 	 +2 \psi -2 \e \Rhot	   
	\right). 
\end{align}

Inserting Eq.~(\ref{eq:absuv}) into Eq.~(\ref{eq:BdGeq}), we get
\begin{align}\label{eq:dotthu}
 	\dot{\theta}_{u}
 	=\,
 	-\frac{\e}{2} + \frac{1}{2} \frac{d \phi}{d t}
 	+ 
 	\frac{
 	 a \left(
 	\Rhot ^2 + 4 \muco \Rhot + 2 \psi - 2 \e \Rhot
 	\right)}
 	{1 - 2\left( 
 		a \Rhot +b\right)},
 \end{align}
 \begin{align}\label{eq:dotthv}
 	\dot{\theta}_{v}
 	=\,
 	\frac{\e}{2} - \frac{1}{2} \frac{d \phi}{d t}
 	+ 
 	\frac{
 	 a \left(
 	\Rhot ^2 + 4 \muco \Rhot + 2 \psi - 2 \e \Rhot
 	\right)}
 	{1 + 2\left( 
 		a \Rhot +b\right)}.
 \end{align}
We define $\e_+$ and $\e_-$ by
\begin{align}\label{eq:epemDEF}
	\e_\pm &\equiv 
    \frac{\Rhot_{\pm}^2 + 4 \muco  \Rhot_{\pm}
  	+2 \psi }{2 \Rhot_{\pm}}.
\end{align}
Eq.~(\ref{eq:dotthu})
[Eq.~(\ref{eq:dotthv})] is valid as long as 
$\e \neq \e_+$ or $t \neq 0$ 
($\e \neq \e_-$ or $t \neq \frac{T}{2}$). 
$\theta_u(\e_+,0)$ and $\theta_v(\e_-,\frac{T}{2})$ 
are unimportant since $|u(\e_+,0)|=0$ and $|v(\e_-,\frac{T}{2})|=0$.

We next switch from the frame rotating with phase $\phi(t) = 2 \masy t + \Phi(t)$ 
to that rotating only with its linear piece $2 \masy t$. 
This is equivalent to replacing $d \phi/ d t \rightarrow 2 \masy$ in 
Eqs.~(\ref{eq:dotthu}) and (\ref{eq:dotthv}).
The resulting equations depend on $t$ only through $\Rhot$, 
which implies that $\theta_{u}$ ($\theta_{v})$ should be a sum 
of a time-periodic function and a linear piece $-E_u t$ ($-E_v t$).
The intercept $E_u$ ($E_v)$ is determined by integrating 
Eq.~(\ref{eq:dotthu}) [Eq.~(\ref{eq:dotthv})] over one period,
 \begin{align}\label{eq:Eu}
 \begin{aligned}
 	E_u(\e)
 	=&\,
 	-\frac{1}{T} 
 	\int_{0}^{T} dt \,
 	\dot{\theta}_{u}(\e,t)
 	\\
	=&\,
 	 \frac{\e}{2}
 	 -\masy
 	 -\frac{1}{T} 
 		\int_{0}^{T}
 		dt \, f_- (\e,t) ,
 		\\
     E_v(\e)
 	=&\,
 	-\frac{1}{T} 
 	\int_{0}^{T} dt \,
 	\dot{\theta}_{v}(\e,t)
 	\\ 
	=&\,
 	 -\frac{\e}{2}
 	 +\masy
 	 -\frac{1}{T} 
 		\int_{0}^{T}
 		dt \, f_+(\e,t).
 \end{aligned}
 \end{align}
 Here
 \begin{align}\label{eq:Ev}
 \begin{aligned}
	f_\pm(\e,t)\equiv
	\frac{a(\e)
 		  \left[ 
 		  \Rhot^2(t) + 4 \muco \Rhot(t) + 2\psi -2 \e \Rhot(t)
 		  \right] }
 		  {1 \pm 2\left[
 		  a(\e) \Rhot(t) + b(\e) \right] }.
 \end{aligned}
 \end{align} 
We find that 
$E_v(\e) - E_u(\e) = 
-
n_0(\e) \Omega$ 
from Eq.~(\ref{eq:thetauv}), where 
$n_0(\e)=\theta(\e - \e_{+})\theta(\e_{-} - \e)$.
As a result,
	$\{
	  u_{0} (\e,t)
	,\,
  	  v_{0} (\e,t)
	\}$
is a Floquet state and can be parameterized as
$	\{
 	 | u_{0}(\e,t) | 
 	 e^{- i \Theta_{u}(\e,t)}
 	 ,\,
 	 | v_{0}(\e,t)| 
 	 e^{- i \Theta_{v}(\e,t)}
 	\}
 	e^{+ i E_0(\e) t }.
$
Here, the quasi-energy 
$E_0 \equiv E_u $(mod $\Omega$), 
$\Theta_{u}(t+T) \equiv \Theta_{u}(t)$ (mod $2\pi$), 
and
$\Theta_{v}(t+T) \equiv \Theta_{v}(t)$ (mod $2\pi$). 
$\Theta_{u},\Theta_{v}$ can be obtained by integrating Eqs.~(\ref{eq:dotthu}) and (\ref{eq:dotthv}). 
Assuming $\Rhot(t=0)=\Rhot_+$ and $\Phi(t=0)=0$, we deduce the integration constant from Eq.~(\ref{eq:thetauv}),
\begin{align}
\begin{aligned}
	\Theta_{u}(\e,0)-\Theta_{v}(\e,0)
	=&\,
	\pi \theta 
	\left( \e-\e_+ \right),
    \\
	\Theta_{u}( \e,\frac{T}{2} )
 	-\Theta_{v}(\e, \frac{T}{2})
  	= & \,
   \pi \theta 
    \left(   \e-\e_- \right),
\end{aligned}
\end{align}
Then we obtain 
 \begin{align}\label{eq:ResTorphase}
 	\Theta_{u} (\e,t)
 	=&\,
 	 \int_{0}^{t}
 	dt' f_-(\e,t')
 	+ \left[ -\frac{\e}{2} + \masy +E_0(\e) \right] t,
	\nonumber\\
 	\Theta_{v} (\e,t)
 	=&\,
	-\pi \theta \left( \e-\e_+ \right)
 	+ \int_{0}^{t}
 	dt' f_+(\e,t')
	\nonumber\\
	&\,
 	+\left[ \frac{\e}{2} - \masy + E_0(\e) \right]t.
 \end{align}
In the scheme where we take the quasienergy $E_0 = E_u$, 
we have
$\Theta_{u}(\e,t+T)=\Theta_{u}(\e,t)$,
$\Theta_{v}(\e,t+T)=\Theta_{v}(\e,t) 
+
2\pi n_0(\e)$, 
and Eq.~(\ref{eq:ResTorphase}) simplifies to
 \begin{align}\label{eq:ResTorphase2}
 	\Theta_{u} (\e,t)
 	=&\,
 	\left[
	\int_{0}^{t}
 	dt'
 	 -\frac{t}{T} 
 		\int_{0}^{T}
 		dt' \,
	\right] 
	f_-(\e,t'),
	\nonumber\\
 	\Theta_{v} (\e,t)
 	=&\,
	\left[
	\int_{0}^{t}
 	dt'
 	 -\frac{t}{T} 
 		\int_{0}^{T}
 		dt'
		\, 
	\right]
	f_+(\e,t')
 		-
 		\pi \theta \left( \e-\e_+ \right)
 	\nonumber\\
 		&
		+	
		2\pi \theta \left( \e-\e_+ \right)\theta\left( \e_--\e \right)
 		\frac{t}{T}.
 \end{align}
Combining Eqs.~(\ref{eq:absuv}) and (\ref{eq:ResTorphase}), we obtain the complete expression 
for $\begin{bmatrix}u_0(\e,t)&v_0(\e,t)\end{bmatrix}^{T}$.
Then the spinor of the 
Lax reduced 
solution can be written as
\begin{align}\label{eq:SpinorResTor}
\begin{aligned}
 	\begin{bmatrix}
 	  u (\e,t)
 	  \\
   	  v (\e,t)
 	\end{bmatrix}
 	= &\,
 	{\textstyle{\sqrt{\frac{1+\zeta(\e)}{2}}}}
 	\begin{bmatrix}
 	 | u_{0} (\e,t) | 
 	 e^{- i \Theta_{u}(\e,t)}
 	 \\
 	 | v_{0}(\e,t) | 
 	 e^{- i \Theta_{v}(\e,t)}
 	\end{bmatrix}
 	e^{+ i E_0(\e) t }
 	\\
 	+&
	{\textstyle{\sqrt{\frac{1-\zeta(\e)}{2}}}}
 	 \begin{bmatrix}
 	 | v_{0}(\e,t) | 
 	 e^{ i \Theta_{v}(\e,t)}
 	 	 \\
 	 -| u_{0}(\e,t) | 
 	 	 e^{ i \Theta_{u}(\e,t)}
 	 	\end{bmatrix}
 	 	e^{- i E_0(\e) t +i \Gamma(\e)}.
 \end{aligned}	 
 \end{align} 
Here $\Gamma(\e)$ is some time-independent phase.

\subsubsection{Initial condition}

We assume that the general solution to the self-consistent BdG 
equations following a quench takes a similar form,
\begin{align}\label{eq:Spinor}
\begin{aligned}
 	\begin{bmatrix}
 	  u (\e,t)
 	  \\
   	  v (\e,t)
 	\end{bmatrix}
 	= &\,
 	{\textstyle{\sqrt{\frac{1-\gamma(\e)}{2}}}}
 	\begin{bmatrix}
 	 | u_{0} (\e,t) | 
 	 e^{- i \Theta_{u}(\e,t)}
 	 \\
 	 | v_{0}(\e,t) | 
 	 e^{- i \Theta_{v}(\e,t)}
 	\end{bmatrix}
 	e^{+ i E_0(\e) t }
 	\\
 	+&
	{\textstyle{\sqrt{\frac{1+\gamma(\e)}{2}}}}
 	 \begin{bmatrix}
 	 | v_{0}(\e,t) | 
 	 e^{ i \Theta_{v}(\e,t)}
 	 	 \\
 	 -| u_{0}(\e,t) | 
 	 	 e^{ i \Theta_{u}(\e,t)}
 	 	\end{bmatrix}
 	 	e^{- i E_0(\e) t +i \Gamma(\e)}.
 \end{aligned}	 
 \end{align}
The Lax reduced solution in Eq.~(\ref{eq:SpinorResTor}) is an extreme example where 
$\gamma=-\zeta$ only takes values equal to $\pm 1$. 
In order to solve the self-consistent conditions in Eq.~(\ref{eq:ResTorCons}), 
$\zeta(\e)$ must exhibit a discontinuity (``Fermi step''). The reduced solution therefore cannot
apply to a quench from a BCS initial state, since the latter initially has a smooth pseudospin
texture in momentum space, and the BCS dynamics do not change this. 
The solution for the quench involves replacing $-\zeta(\e)$ with a smooth
$\gamma(\e)$ that nevertheless winds from $+1$ to $-1$ with increasing $\e$. 

The conservation of energy $E$, $z$-component angular momentum J (particle number), and the 
Lax norm allows us to determine the $\gamma_i$ for the post-quench state. 
Substituting Eq.~(\ref{eq:Spinor}) into Eq.~(\ref{eq:spin2coh}), we have
\begin{align}\label{eq:smt}
	s_{i}^{-} (t) 
	=&\, 
	-\gamma_i	
	u^*_{0,i}  v_{0,i}
	\nonumber\\
	+&\frac{\sqrt{1-\gamma_i^2}}{2}
	\left(
	  -e^{i\Gamma_i}
	  u^{*2}_{0,i}  
	  e^{- 2 i E_i t}
	  +e^{-i\Gamma_i}
	  v_{0,i} ^2 
	  e^{ 2 i E_i t}
	\right) ,
	\nonumber\\
	s_{i}^{z} (t) 
	=&\, -\gamma_i	
	\frac{1}{2}
	(| v_{0,i} (t)|^2- | u_{0,i} (t)|^2)
	\nonumber\\
	+&\frac{
	 \sqrt{1-\gamma_i^2}}
	 {2}
	\left(
	\begin{aligned}
	&\,
	  -e^{-i\Gamma_i}
	  u_{0,i} v_{0,i}   
	  e^{ 2 i E_i t}
	\\&\,
	  -e^{i\Gamma_i}
	  u^*_{0,i} v^*_{0,i} 
	  e^{ -2 i E_i t}
	\end{aligned}
	\right).
\end{align}
These can be considered as a combination of $\vec{s}_{0,i}(t)$ and an oscillating spin with 
energy-dependent frequency $\vec{s}_{\mathsf{osc}, i}(t)$, weighted by $-\gamma_i$ and $\sqrt{1-\gamma_i^2}$ separately, i.e.
$	\vec{s}_i (t)=
	-\gamma_i \vec{s}_{0,i} 
	+\sqrt{1-\gamma_i^2} \,
	 \vec{s}_{\mathsf{osc}, i}$.
As $t \rightarrow \infty$, through repeated integration-by-parts or the saddle point approximation, we 
find that the contribution of $\vec{s}_{\mathsf{osc}, i}$ to $E$, $J$, and the Lax norm vanishes. Therefore, 
all of the conserved quantities can be obtained by replacing $\zeta$ with $-\gamma$ in 
Eqs.~(\ref{eq:ResTorCons})--(\ref{eq:ResTorL2}), which leads to
\begin{align}\label{eq:IIIJE}
\begin{aligned}
	& \sum_i \gamma_i a_i=\,0 
	,&&
	\sum_i \gamma_i \varepsilon_i a_i=\,-\frac{1}{G},
	\\
	&\sum_{i} 
	\gamma_i 
     b_{i} = -\,J
     ,&&
	\sum_{i} 
	\gamma_i
	\varepsilon_i b_{i} 
	=-\,E,
\end{aligned}
\end{align}
and
\begin{align}\label{eq:IIIL2}
\begin{aligned}
	L_2 (u) 
	=&\,
	\left( 
	  \sum_j 
	   \frac{\gamma_j \e_j }
	     {2 \sqrt{ \mathcal{Q}_4(\e_j) }}
	     \frac{1}{\e_j - u}
	     \right) ^2
	     \mathcal{Q}_4(u).
\end{aligned}
\end{align}
We now let $u$ in Eq.~(\ref{eq:IIIL2}) approach the positive real axis from above and below, 
$u \rightarrow u \pm i \eta$, $u \in \mathbb{R^+}$ and $\eta \rightarrow 0$. 
From the residue theorem,
\begin{align}
\begin{aligned}
	L_2 ( u \pm i \eta ) 
	=\,
	\nu^2
	\Bigg[
	P
	\intem
	   \frac{\gamma (\e) \e }
	    {2 \sqrt{ \mathcal{Q}_4(\e) }}
	   \frac{1}{\e - u}
	   \\
	  \pm
	  i \pi
	  \frac{\gamma(u) u}
	    {2 \sqrt{\mathcal{Q}_4(u)} }
	\Bigg]^2
	 \mathcal{Q}_4(u),	
\end{aligned}
\end{align}
where $\nu \equiv \mathcal{L}^2/8\pi$ is the bare density of states, and $\mathcal{L}$ denotes the linear system size. 
This is equal to the Lax norm of the pre-quench state, 
\[
	L_2 ( u \pm i \eta ) =\, \frac{ \nu^2 }{4} I_\mp(u),
\]
which gives
\begin{align}\label{eq:IIIgamma}
	|\gamma (\e)|
	=\,
	\frac{ |\sqrt{I_+( \e )}
	   -\sqrt{I_-( \e )}|}
	 {2 \pi \e}.
\end{align}
This shares the same form as the pseudospin distribution function in phases I and II. 
Here, $I_{+}(\e)$ is a certain function that depends upon the parameters of the pre-quench state,
as well as the strength and direction of the quench (i.e., the quench coordinates $\{\Di,\Df\}$),
while $I_{-}(\e)$ is its complex conjugate;
an explicit expression appears in \cite{LONG}.
The \emph{sign} of $\gamma$ is determined by 
enforcing continuity whenever $\gamma(\e) \rightarrow 0$ with a non-zero slope \cite{LONG}.
We find that $\gamma(\e)$ winds exactly once from $+1$ at small $\e$ to $-1$ at large $\e$ for
all quenches in phase III; see e.g.\ Fig.~\ref{fig:DF}.
For all phase III quenches that we have investigated including ``A'' and ``B'' indicated in 
Fig.~\ref{fig:diagram}, we have used the result for $\gamma(\e)$ and verified that all four
relations in Eq.~(\ref{eq:IIIJE}) hold.

We find that $ \Gamma(\e) $ in Eq.~(\ref{eq:Spinor}) must be $0$ or $\pi$ due to the 
invariance of the pseudospin equations of motion under the effective time-reversal transformation
\cite{LONG}
 \begin{align}
 \begin{aligned}
 	s^z (\e,t) 
 	&\rightarrow
 	s^z (\e,-t),
 	\\
 	s^{\pm} (\e,t) 
 	&\rightarrow
 	s^{\mp}(\e,-t),
 	\\
 	\Delta (t) 
 	&\rightarrow
 	\Delta^*(-t).		 
 \end{aligned}
 \end{align}

 \subsubsection{Phase II,III border}

On the border of phases II and III,  $\Delta_1$ approaches $0$, and the time evolution of
the modulus and phase of the order parameter is given by 
 \begin{align}\label{eq:borderDelta}
 \begin{aligned}
 	\sqrt{\Rhot (t)}
 	=&\,
 	\Delta_2 
	\quad
	\text{(const.)}, 
 	\\
 	\phi (t)
 	=&\,
 	2 \mu _2 t.
 \end{aligned}
 \end{align}
 The Lax reduced solution on the border can be obtained from Eqs.~(\ref{eq:borderDelta}), (\ref{eq:ResTor}) and (\ref{eq:abDef}),
 \begin{align}\label{eq:BorderResTor}
 \begin{aligned}
 	s^z(\e)
 	=&\,
 	-
	\zeta (\e)
 	\sgn (\e-2\mu_1)
 	\frac{(\frac{\e}{2}-\mu_2)}
 	{2 E_2 (\e)}
 	\\
 	s^-(\e)
 	=&\,	
	-
 	\zeta (\e)
 	\sgn (\e-2\mu_1)
 	\frac{\sqrt{\e} \Delta_2}
 		{2 E_2 (\e)}
 		e^{-2i \mu_2 t}.
 \end{aligned}
 \end{align}     
These are the pseudospin configurations for ground (excited) state Cooper pairs 
when $\zeta(\e)\sgn(\e-2\mu_1)=1(-1)$.

In addition, following all the equations for the spinor in phase III, we find
\begin{align}
\begin{aligned}
	E_0(\e) & =\,\sgn(\e-2\mu_1)E_2(\e),
	\\
		|u_0(\e)|
		&=\,
		\sqrt{
		  \frac{1}{2}+
		   \frac{\sgn{(\e - 2 \mu_1)}(\frac{\e}{2}-\mu_2)}
		   {2 E_2 (\e)}
		   },
		\\
		|v_0(\e)|
		&=\,
		\sqrt{
		\frac{1}{2}-
		  \frac{\sgn{(\e -2 \mu_1)}(\frac{\e}{2}-\mu_2)}
		 {2 E_2 (\e)}
		  },
		  \\
	\Theta_{u} (\e)
	&=\, 0 ,\quad
	\Theta_{v} (\e)
      =\,
	 -\pi \theta (\e -2 \mu_1),
\end{aligned}
\end{align}
which is consistent with pseudospin configuration [Eq.~(\ref{eq:BorderResTor})]. 
Since $\e_+=\e_-=2 \mu_1 $, $E_u=E_v$ holds for any value of $\e$. 
It is obvious that Eq.~(\ref{eq:Spinor}) will take the form of the phase II 
wavefunction after applying the transformation $\gamma \rightarrow \sgn(\e-2 \mu_1)\gamma$.


\begin{thebibliography}{99}
\bibitem{OkaAoki09}
	T. Oka and H. Aoki,
	Phys. Rev. B {\bf 79}, 081406(R) (2009).
\bibitem{Lindner11}
	N. H. Lindner, G. Refael, and V. Galitski, 
	Nat. Phys. {\bf 7}, 490 (2011).
\bibitem{Kitagawa11}
	T. Kitagawa, T. Oka, A. Brataas, L. Fu, and E. Demler, 
	Phys. Rev. B {\bf 84}, 235108 (2011).
\bibitem{Gu11}
	Z. Gu, H. A. Fertig, D. P. Arovas, and A. Auerbach, 
	Phys. Rev. Lett. {\bf 107}, 216601 (2011).
\bibitem{Rudner13}
	M. S. Rudner, N. H. Lindner, E. Berg, and M. Levin,
	Phys. Rev. X {\bf 3}, 031005 (2013).
\bibitem{Kundu14}
	A. Kundu, H. A. Fertig, and B. Seradjeh, 
	Phys. Rev. Lett. {\bf 113}, 236803 (2014).
\bibitem{Usaj14}
	L. E. F. Foa Torres, P. M. Perez-Piskunow, C. A. Balseiro, and G. Usaj,
	Phys. Rev. Lett. {\bf 113}, 266801 (2014).
\bibitem{SHORT}
	M. S. Foster, V. Gurarie, M. Dzero, and E. A. Yuzbashyan,
	Phys. Rev. Lett. {\bf 113}, 076403 (2014).
\bibitem{LONG}
	M. S. Foster, M. Dzero, V. Gurarie, and E. A. Yuzbashyan,
	Phys. Rev. B {\bf 88}, 104511 (2013).	
\bibitem{DongPu14}
	Y. Dong, L. Dong, M. Gong, and H. Pu,
	Nat. Comm. {\bf 6}, 6103 (2014).
\bibitem{S-wave1}
	R. A. Barankov, L. S. Levitov, and B. Z. Spivak,
	Phys. Rev. Lett. \textbf{93}, 160401 (2004).
\bibitem{YuzbashyanAltshuler06}
	E. A. Yuzbashyan, O. Tsyplyatyev, and B. L. Altshuler,
	Phys. Rev. Lett. \textbf{96}, 097005 (2006).
\bibitem{BarankovLevitov06}
	R. A. Barankov and L. S. Levitov,
	Phys. Rev. Lett. \textbf{96}, 230403 (2006).
\bibitem{BarankovLevitov06-II}
	R. A. Barankov and L. S. Levitov,
	Phys. Rev. A \textbf{73}, 033614 (2006).
\bibitem{Chamon13}
	T. Iadecola, D. Campbell, C. Chamon, C.-Y. Hou, R. Jackiw, S.-Y. Pi, and S. V. Kusminskiy,
	Phys. Rev. Lett. {\bf 110}, 176603 (2013).
\bibitem{Mitra14}
	H. Dehghani, T. Oka, and A. Mitra,
	Phys. Rev. B {\bf 90}, 195429 (2014); 
	Phys. Rev. B, {\bf 91}, 155422 (2015).
\bibitem{DAlessioRigol14}	
	L. D'Alessio and M. Rigol,
	Nat. Comm. {\bf 6}, 8336 (2015).
\bibitem{rfswave}
	M. Dzero, E. A. Yuzbashyan, B. L. Altshuler, and P. Coleman,
	Phys. Rev. Lett. {\bf 99}, 160402 (2007).	
\bibitem{Schrieffer}
	J. R. Schrieffer,
	\textit{Theory of Superconductivity}
	(Perseus Books, Reading, Massachusettes, 1983).
\bibitem{GedikARPES}
	Y. H. Wang, H. Steinberg, P. Jarillo-Herrero, and N. Gedik, 
	Science {\bf 342}, 453 (2013).
\bibitem{Shimano1}
	R. Matsunaga, Y. I. Hamada, K. Makise, Y. Uzawa, H. Terai, Z. Wang, and R. Shimano,
	Phys. Rev. Lett. \textbf{111}, 057002 (2013).
\bibitem{Shimano2}
	R. Matsunaga, N. Tsuji, H. Fujita, A. Sugioka, K. Makise, Y. Uzawa, H. Terai, 
	Z. Wang, H. Aoki, and R. Shimano,
	Science {\bf 345}, 1145 (2014). 
\bibitem{rf-Tun-SecondDist}
	There is a second important difference between tunneling and rf, besides the distribution function
	dependence of the latter. Because the rf probe is global,  
	it conserves momentum.
	This leads to differences in terms of where gaps 
	appear in the spectrum \cite{rfpwaveEq}; see Secs.~\ref{Sec:IIrf} and \ref{Sec:IITun} for more details. 
\bibitem{rfpwaveEq}
	E. Grosfeld, N. R. Cooper, A. Stern, and R. Ilan, 
	Phys. Rev. B {\bf 76}, 104516 (2007).
\bibitem{Gurarie11}
	A. M. Essin and V. Gurarie, 
	Phys. Rev. B {\bf 84}, 125132 (2011).
\bibitem{Tsuji2011}
		N. Tsuji, T. Oka, P. Werner, and H. Aoki,
		Phys. Rev. Lett. \textbf{106}, 236401 (2011).
\bibitem{DzeroYuzbashyan06}
	E. A. Yuzbashyan and M. Dzero,
	Phys. Rev. Lett. \textbf{96}, 230404 (2006).
\bibitem{Caio15}
	M. D. Caio, N. R. Cooper, and M. J. Bhaseen, 
	arXiv:1504.01910 [cond-mat.str-el]
\bibitem{YuzbashLax1}
	E. A. Yuzbashyan, B. L. Altshuler, V. B. Kuznetsov, and V. Z. Enolskii,
	Phys. Rev. B \textbf{72}, 220503(R) (2005).
\bibitem{YuzbashLax2}
	E. A. Yuzbashyan, B. L. Altshuler, V. B. Kuznetsov, and V. Z. Enolskii,
	J. Phys. A \textbf{38}, 7831 (2005).
\bibitem{footnote--Phase Diagram}
	This phase diagram was computed in \cite{LONG}. 
	The natural scale for $\Delta$ is set by the equilibrium transition $\Dqcp$;
	the value of $(\Dqcp)^2$ is determined by the fixed particle density $n$ 
	and the condition
	$\mu = 0$, up to a logarithmic dependence upon an ultraviolet cutoff $\Lambda$.
	To generate Fig.~\ref{fig:diagram}, we set $\Lambda = 50 \mathcal{E}_F = 100 \pi n$. 
	The same assumption is made throughout the paper to evaluate parameters
	of the dynamics, see Ref.~\cite{LONG} for details.
\bibitem{footnote--p+/-ip}
	Relative to the $s$-wave case, there is an additional 	
	complication for quenches in our 2D $p$-wave superfluid,
	if the initial state is a non-interacting Fermi liquid.
	In fact, for both $s$- and $p$-wave cases, the Fermi step is 
	a state of metastable equilibrium at the level of self-consistent
	mean field theory, and so does not evolve in time. In reality,
	quantum or thermal fluctuations will induce the instability of
	the normal state. For such a quench in the $p$-wave case,
	one would expect both $p+ip$ and $p-ip$ fluctuations;
	the subsequent evolution must track the interplay of 
	both pairing channels \cite{SHORT}. We 
	expect that the reduced BCS model retaining both 
	channels is not integrable, so that the Lax technology
	used to solve the case of pure $p+ip$ order is not
	immediately applicable. 
	This problem deserves further consideration. 
\bibitem{Read2000}
	N. Read and D. Green,
	Phys. Rev. B \textbf{61}, 10267 (2000).
\bibitem{VolovikBook}
	G. E. Volovik,
	\emph{The Universe in a {Helium} Droplet}
	(Oxford University Press, Oxford, 2003).	
\end{thebibliography}
\end{document}